%% file: main.tex
\documentclass{article}

\usepackage{arxiv}

\usepackage{enumitem}
\usepackage{amssymb}
\usepackage{xcolor}
\usepackage{adjustbox}
\usepackage{graphicx} 
\usepackage{hyperref}
\usepackage{caption,subcaption}
\usepackage[linesnumbered,ruled,vlined]{algorithm2e}
\usepackage{listings}
\usepackage{multirow}
\usepackage{colortbl}
\usepackage{booktabs}
\usepackage{hhline}
\usepackage{appendix}
\usepackage{pifont}
\usepackage{tikz}

\usepackage{threeparttable}
\usepackage{tcolorbox}
\usepackage[affil-it]{authblk}

\newcommand{\sysname}{\textsc{InsDet}\xspace}

\newcommand{\tab}{\hspace*{1em}}

\newcommand{\code}[1]{{\fontfamily{cmtt}\fontseries{m}\fontshape{n}\selectfont\small{#1}}}

\newcommand{\tool}{\textsc{InsDet}\xspace}

\newsavebox{\mybox}

\definecolor{mygreen}{RGB}{0 ,205, 102}
\begin{document}

\title{Automatically Locating  ARM Instructions Deviation between Real Devices and CPU Emulators}

	\author[1,2]{Muhui Jiang}
	\author[1]{Tianyi Xu}
	\author[1]{Yajin Zhou\thanks{Corresponding author (yajin\_zhou@zju.edu.cn).}\hspace*{0.4em}}
	\author[1]{Yufeng Hu}
	\author[1]{Ming Zhong}
	\author[1]{Lei Wu}
	\author[2]{Xiapu Luo}
	\author[1]{Kui Ren}
	\affil[1]{Zhejiang University}
	\affil[2]{The Hong Kong Polytechnic
		University}

\maketitle

\begin{abstract}
Emulator is widely used to build dynamic analysis frameworks due to its fine-grained tracing capability, full system monitoring functionality, and scalability of running on different operating systems and architectures. However, whether the emulator is consistent with real devices is unknown. To understand this problem, we aim to automatically locate inconsistent instructions, which behave differently between emulators and real devices. 

We target ARM architecture, which provides machine readable specification. Based on the specification, we propose a test case generator by  designing and implementing the first symbolic execution engine for ARM architecture specification language (ASL). We generate 2,774,649 representative instruction streams and conduct differential testing with these instruction streams between four ARM real devices in different architecture versions (i.e., ARMv5, ARMv6, ARMv7-a, and ARMv8-a) and the state-of-the-art emulators (i.e., QEMU). We locate 155,642 inconsistent instruction streams, which cover 30\% of all instruction encodings and 47.8\% of the instructions. We find undefined implementation in ARM manual and implementation bugs of QEMU are the major causes of inconsistencies.
Furthermore, we discover four QEMU bugs, which are confirmed and patched by the developers, covering 13 instruction encodings including the most commonly used ones (e.g., \code{STR}, \code{BLX}). With the inconsistent instructions, we build three security applications and demonstrate the capability of these instructions on detecting emulators, anti-emulation, and anti-fuzzing. 

\end{abstract}

\input{01_introduction.tex}

\input{02_background.tex}

\input{03_design.tex}

\input{04_implementation.tex}

\input{05_evaluation.tex}

\input{06_discussion.tex}

\input{07_related_work.tex}

\input{08_conclusion.tex}

\bibliographystyle{plain}
\bibliography{main}



\end{document}

%% file: 01_introduction.tex
\section{Introduction}
CPU emulator is a powerful tool as it provides fundamental functionalities (e.g., tracing, record and replay) for the dynamic analysis. Though hardware-based tracing techniques 
exist, they have limitations compared with software emulation. For example, ARM ETM has limited Embedded Trace Buffer (ETB). The size of ETB of the Juno Development Board is 64KB~\footnote{The ETB size of different SoCs may be different. However, it's usually limited due to the chip cost and size.}~\cite{Juno}. 
On the contrary, software emulation 
is capable of tracing the whole program, provides user-friendly APIs for runtime instrumentation, and can run on multiple operating systems (e.g., Windows and Linux) and host machines in different architectures.
Nevertheless, software emulation complements the hardware-based tracing and provides rich functionalities that dynamic analysis systems can build upon.   

Indeed, 
many dynamic analysis frameworks~\cite{davanian2019decaf++,henderson2014decaf,yan2012droidscope,alwabel2014safe,wei2015mose,carmony2016extract, chipounov2011s2e,jiang2010stealthy, chen2016firmadyne,feng2014mace,luo2016repackage,kim2020firmae,johnson2021jetset} are built based on the state-of-the-art CPU emulator, i.e., QEMU, to conduct malware analysis, live-patching, crash analysis and etc.
Meanwhile, many fuzzing tools utilize CPU emulators to fuzz
binaries, e.g., the QEMU mode of AFL~\cite{afl_qemu}, FirmAFL~\cite{zheng2019firm}, P2IM~\cite{feng2019p2im}, HALucinator~\cite{clements2019hal} and TriforceAFL~\cite{TriforceAFL}.

The wide adoption of software emulation usually has an implicit assumption that 
the execution result of an instruction on the CPU emulator and the real device is identical, thus running a program on the CPU emulator can reflect the result on the real hardware. 
\textit{However, whether this assumption really holds in reality is unknown.}
In fact, the execution result could be different (as shown in our work),
either because the CPU emulator has bugs or because it uses a different implementation from the real device. These differences impede the reliability of emulator-based dynamic analysis. For instance, the malware can abuse the differences to protect the malicious behaviors from being analyzed in the emulator~\cite{raffetseder2007detecting,jang2019rethinking,issa2012anti,lictua2018anti}.

In this work, we aim to automatically locate inconsistent instructions between  real devices and the CPU emulator for the ARM architecture. If an instruction behaves differently between them, then it is an inconsistent instruction. Although previous research~\cite{lorenzo2012,martignoni2010testing,martignoni2009testing,martignoni2013methodology}
provides valuable insights, they are limited to the x86/x64
architecture and cannot be directly applied to the ARM architecture.
Our work leverages the differential testing~\cite{mckeeman1998differential} for the purpose. Specifically, we provide the same instruction stream~\footnote{In this paper, instruction and instruction stream represent different meanings. For example , we call STR (immediate) an instruction.  We call the concrete bytecode (i.e., 0xf84f0ddd) an instruction stream. See Section~\ref{sec:term}} to both the real device and a CPU emulator, and compare the execution result to check whether it is an inconsistent one.

Though the basic idea is straightforward, it faces the following two challenges.
\textit{First}, the ARM architecture has multiple versions (e.g., ARM v5, v6, v7 and v8), different register widths (16 bits or 32 bits) and instruction sets. Besides, it has mixed instruction modes (ARM, Thumb-1 and Thumb-2). 
Thus, how to generate effective test cases, i.e. instruction streams that cover previously mentioned architecture variants, while at the same time generating only  necessary test cases to save the time cost, is the first challenge. 
Notice that if we naively enumerate 32-bit instruction streams, the number of test cases would be $2^{32}$, which is inefficient, if not possible, to be evaluated. Meanwhile, randomly generated instruction streams are not representative and many instructions are not covered (Section~\ref{sec:rq1}).
\textit{Second}, for each test case, we should provide a deterministic
environment to execute the single instruction stream and automatically compare the result after
the execution. This requires us to set up the same context (with CPU registers
and memory regions) before the execution and compare
the context afterwards.

Our system solves the challenges with the following two key techniques.

\noindent\textbf{Syntax and semantics aware test case generation}\tab
To generate representative instruction streams, we propose a
syntax and semantics aware test case generation methodology.
Each ARM instruction consists of several 
\textit{encoding schemas}, which is called \textit{instruction encodings} in this paper, that define the instruction's structure (syntax). Each encoding schema maps to one decoding and execution logic that defines
the instruction semantics.
The encoding schema shows which parts of an instruction are
constants and which parts can be mutated (Figure~\ref{fig:motivation}(a)).
The non-constant parts of an instruction are called
\textit{encoding symbols} in this paper.
The decoding and execution logic is
expressed in the ARM's Architecture Specific Language (ASL)~\cite{reid2016trustworthy}
. We call it the \textit{ASL code} in this paper
(Figure~\ref{fig:motivation}(b) and (c)).
The ASL code executes based on the concrete values of the encoding symbols. 
For instance, if the concrete value of the encoding
symbol \code{W} (the eighth bit of \code{STR (immediate)} instruction) is \code{1}, then
the new address will be written back into the destination register \code{Rn} (line 4 of Figure~\ref{fig:motivation}(c)).

Specifically, during the test case generation,
we first take the syntax-aware strategy. For each encoding symbol, we mutate it based on
pre-defined rules. For instance, for the immediate value symbol,
the values in the mutation set cover the maximum value, the
minimum value and a fixed number of random values. This strategy
generates syntactically correct instructions.

We further take a semantics-aware strategy to generate more
instruction streams. That's because the previous strategy may only
cover limited  instruction semantics as different encoding symbol values can result in different decoding and executing behaviors (Section~\ref{subsec:motivation_example}).
To this end, we extract the constraints in ASL code of decoding and executing.
We solve the constraints and their negations by designing and implementing the first symbolic execution engine for ASL to find the  satisfied
values of the encoding symbols. By doing
so, the generated test cases can cover different semantics
of an instruction. 

\noindent\textbf{Deterministic differential testing engine}\tab
Our differential testing engine uses the generated test cases
as inputs. To get a deterministic testing result, we provide
the same context when executing an instruction
stream on a
real CPU and an emulator. Besides, an instruction stream cannot
be directly loaded and executed by the emulator, we 
carefully design a template binary that converts one instruction stream to
a testing binary by inserting the prologue and epilogue instructions. The prologue instructions aim to set the execution environment while the epilogue instructions will dump the execution result for comparison to check whether the testing instruction stream is an inconsistent one.

We have implemented a prototype system called \tool.
Our test case generator generated
$2,774,649$ instruction streams that cover all the $1,998$
ARM instruction encodings from $1,070$ instructions in four instruction sets (i.e., A64, A32, T32, and T16).

On the contrary, the same number of randomly generated instruction streams can only cover $51.4\%$ Instructions.
This result shows the sufficiency of our test case generator.

We then feed these test cases into our differential testing engine. By comparing the result
between the state-of-the-art emulator (i.e., QEMU) and real devices with four architecture versions (ARMv5, ARMv6, ARMv7-a, and ARMv8-a), our system detected $155,642$ inconsistent instruction streams. Furthermore, these inconsistent instruction streams cover $47.8\%$ of the instructions.

We then explore the root causes of them. It turns out that implementation bugs
of QEMU and the undefined implementation in the ARM manual (i.e., the instruction does not have a well-defined behavior) are the major causes. We discovered four implementation bugs of QEMU and all of
them have been confirmed by developers. These bugs influence $13$ instruction encodings,
including commonly used instructions, e.g., \code{BLX}, \code{STR}. 

To show the usage of our findings, we further build three applications, i.e., emulator detection, anti-emulation and anti-fuzzing. By (ab)using inconsistent instructions, a program can successfully detect the existence of the CPU emulator and prevent the malicious behavior from being monitored by the dynamic analysis framework based on QEMU. Besides, the coverage of the program being fuzzed inside an emulator can be highly decreased. Note that, we only use these applications to demonstrate the usage scenarios of our findings. There may exist other applications, and we do not claim the contribution of them in this paper.

Our work makes the following main contributions.

\noindent\textbf{New test case generator}\tab We propose a test case generator by introducing the first symbolic execution engine for
ARM ASL code. It can generate representative instruction streams that sufficiently cover different instructions (encodings) and semantics.

\noindent\textbf{New prototype system}\tab We implement
a prototype  system named \sysname that consists of a test case generator and a differential testing engine. Our experiments showed \sysname can automatically locate inconsistent instructions. 

\noindent\textbf{New findings}\tab We explore and report the root cause of the inconsistent instructions. Implementation bugs of QEMU and undefined implementation in ARM manual are the major causes. Furthermore,
four bugs have been discovered and confirmed by QEMU developers. Some of them influence commonly used instructions (e.g., \code{STR}, \code{BLX}).

We will release generated test cases and the source code of our system to engage the community.

%% file: 02_background.tex
\section{Background}
\subsection{Terms}
\label{sec:term}
For better illustration and avoid the potential confusion. We give detailed definition towards the following terms used in this paper. 

\noindent\textbf{Instruction}\tab Instruction denotes the category of ARM instructions in terms of functionality, which is usually represented by its name in ARM manual. For example, \code{STR (immediate)} is an instruction, which aims to store a word from a register to memory.

\noindent\textbf{Instruction Encoding}\tab Instruction encoding refers to the encoding schemas for each instruction. We also call it encoding diagram in this paper. 
One instruction can have several encoding schemas. 

\noindent\textbf{Instruction Stream}\tab Instruction stream refers to the bytecode of an instruction. For example, 0xf84f0ddd, which meets one of the encoding schema of instruction \code{STR (immediate)}. We call 0xf84f0ddd an instruction stream.

\subsection{ARM Instruction and Instruction Encoding}
\label{sec:asl}
Processor specification is important as it can verify the implementation of hardware, compilers, emulators, etc. To formalize the specification, ARM introduced the Architecture specification language (ASL)~\cite{reid2016trustworthy}, which is machine-readable and executable. 

ARM instructions usually have a fixed length (16 bits or 32 bits). According to ARM manual, one instruction may consist of several different instruction encodings, which describe the instruction structure (syntax). Our system generates the instruction streams  that cover all the
instruction encodings (which cover all instructions.)  
Specifically, the instruction encoding describes which parts of the instruction are constant and which
parts are not. 
Each instruction encoding is further described with specific decoding and executing logic.
The decoding and executing logic (expressed in ASL) defines the semantics of the instruction.

\subsection{Instruction Decoding in QEMU}
QEMU is the state-of-the-art CPU emulator that supports multiple CPU architectures.
When executing an instruction stream, it needs to decode the instruction stream.
QEMU adopts a two-stage decoding schema. In the first stage, it matches an instruction
stream with pre-defined patterns, each of them represent multiple instructions. 
Then it distinguishes each instruction encoding based on the concrete value
of the instruction. For instance, QEMU groups
\code{VLD4}, \code{VLD3}, \code{VLD2}, and \code{VLD1} instruction into one group (with one
common pattern) and then identifies them inside the instruction decoding routine.
If no instruction pattern can be found or further decoding routine cannot recognize
an instruction stream, the \code{SIGILL} signal will be raised for the user mode emulation
of QEMU.

%% file: 03_design.tex
\section{Design and Implementation}

\begin{figure}[t]
	\centering
	\includegraphics[width=0.8\linewidth]{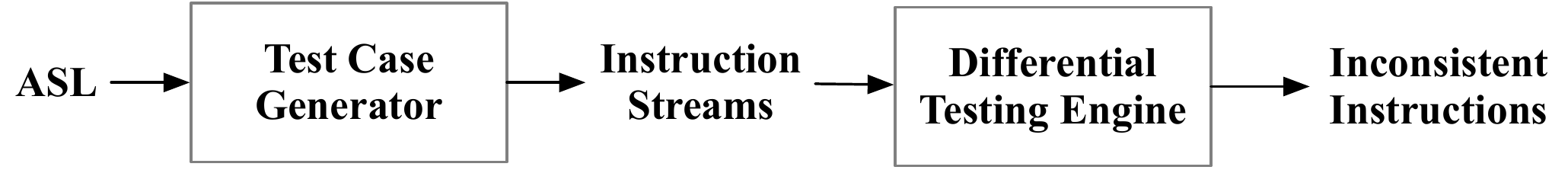}
	
	\caption{\small{The work flow of our system}}

	\label{fig:flow_new}
\end{figure}

Figure~\ref{fig:flow_new} shows the workflow of \sysname,
which consists of a test case generator and a differential testing engine.
First, the test case generator retrieves the ASL code to generate the test
cases (Section~\ref{subsec:case_generator}).
Then, the differential testing engine receives the generated test cases and conducts differential testing between the state-of-the-art emulator (i.e., QEMU) and real devices (Section~\ref{subsec:engine}). The instructions that can result in different behaviors are reported as inconsistent instructions. We further analyze the identified inconsistent instructions
to understand the root cause of them and how they can be (ab)used.

In the following, we first use an inconsistent instruction detected by our system as
a motivating example (Section~\ref{subsec:motivation_example}), and then elaborate the test case generator
and the differential testing engine in Section~\ref{subsec:case_generator} and
Section~\ref{subsec:engine}, respectively.

\subsection{A Motivating Example}
\label{subsec:motivation_example}
\begin{figure}[t]
    \centering
    \begin{subfigure}[b]{\linewidth}
        \centering
        \includegraphics[width=0.9\textwidth]{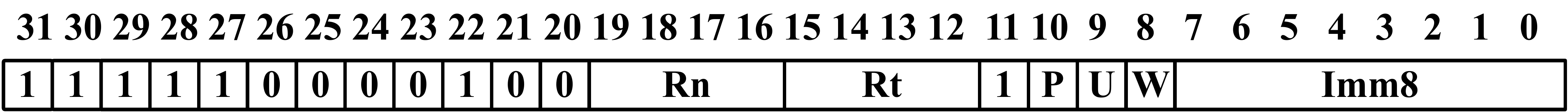}
\caption[Caption for LOF]{\small{The encoding schema of the STR (immediate) instruction in Thumb-2 mode.}}
\label{fig:motivation_encoding}
    \end{subfigure}
\begin{subfigure}[b]{\linewidth}
	\centering
	\begin{lstlisting}
    if Rn == '1111' || (P == '0' && W == '0') then UNDEFINED;
    t = UInt(Rt);  
    n = UInt(Rn);  
    imm32 = ZeroExtend(imm8, 32);
    index = (P == '1');  
    add = (U == '1');  
    wback = (W == '1');
    if t == 15 || (wback && n == t) then UNPREDICTABLE;
	\end{lstlisting}
  
	\caption[Caption for LOF]{\small{The ASL code for decoding the instruction.}}
	\label{fig:motivation_decoding}
\end{subfigure}

\begin{subfigure}[b]{\linewidth}
	\centering
	\begin{lstlisting}
    offset_addr = if add then (R[n] + imm32) else (R[n] - imm32);
    address = if index then offset_addr else R[n];
    MemU[address,4] = R[t];
    if wback then R[n] = offset_addr;
	\end{lstlisting}

	\caption[Caption for LOF]{\small{The ASL code for executing the instruction.}}
	\label{fig:motivation_executing}
\end{subfigure}

\caption{\small{A motivating example.}}

\label{fig:motivation}
\end{figure}

\subsubsection{The encoding schema and semantics of the STR (immediate) instruction}
Figure~\ref{fig:motivation} shows one of the encoding schema of instruction \code{STR (immediate)} and the corresponding ASL code for decoding and
execution logic.
According to the encoding schema in Figure~\ref{fig:motivation_encoding},
the value is constant (i.e., \code{111110000100}) for offset [31:20].
The encoding symbol \code{Rn} and \code{Rt} represent the addressing
register and the source register, respectively. 
The last eight bits ([7:0]) represents a symbol value named \code{imm8}
that will be used as the offset.

Figure~\ref{fig:motivation_decoding} 
shows the ASL code of the decoding logic for the encoding schema.
Note that the ASL code is simplified for presentation.
The complete code can be found in ARM official site~\cite{arm_asl}.

\begin{itemize} [leftmargin=*]
	\item The ASL code at Line 1 checks the value of \code{Rn}, \code{P}, and \code{W}.
	      If the conditions are satisfied (or constraints are met), the instruction stream will be
	      treated as an \code{UNDEFINED} one. Consequently, a \code{SIGILL} signal will be raised in
	      QEMU user mode emulation when an \code{UNDEFINED} instruction stream is executed.
	
	\item  In line 2 and 3, the symbol \code{Rt} and \code{Rn} will be converted to unsigned integer
	\code{t} and \code{n}, respectively. Similarly, the symbol \code{imm8} will be extended into a 32-bit
	integer \code{imm32}. In line 5, 6, and 7, symbol \code{index}, \code{add}, and \code{wback} will be assigned
	according to the value of \code{P}, \code{U}, and \code{W}, respectively.
	
	\item  In line 8, the symbol \code{t}, \code{wback}, and \code{n} will be checked.
	If the constraint of each condition is met, the instruction stream should be treated as
	an \code{UNPREDICTABLE} one. According to ARM's manual, the behavior of an \code{UNPREDICTABLE} instruction stream
	is not defined. The CPU processor vendors and the emulator developers can choose an implementation that
	they think it's proper.
	
\end{itemize}

Similarly, Figure~\ref{fig:motivation_executing} shows the ASL code for the execution logic of the instruction.
The ASL code in Figure~\ref{fig:motivation_decoding} and Figure~\ref{fig:motivation_executing} defines the 
semantics of the instruction.

\subsubsection{Test case generation}
By analyzing the encoding schema, \sysname generates the test cases by mutating the non-constant fields,
including \code{Rn}, \code{Rt}, \code{P}, \code{U}, \code{W} and \code{Imm8}. This can generate
syntactically correct instructions. However, this step is not enough, since it may not
generate the values that satisfy the symbolic expression in the ASL code. For instance,
one symbolic expression in line 8 of Figure~\ref{fig:motivation_decoding} is
\code{t == 15}. The random values generated in the first step may not satisfy this expression
(all of them are not equal to 15). To this end, we leverage a constraint
solver to find the concrete value of the encoding symbol \code{Rt} that satisfies the constraint, i.e.,
\code{15}. Note that, we only use this to illustrate the basic idea. The concrete value
\code{15} of \code{Rt} likely has been generated in the first step.
We take similar actions to solve the constraints for other symbols in line 1 (\code{add}),
2 (\code{index}) and 4 (\code{wback}) of Figure~\ref{fig:motivation_executing}.
During this process, we generated $576$ instruction streams as test cases in total.

\subsubsection{Differential testing}
We feed each instruction stream into our differential testing engine.
The engine generates a corresponding ELF binary for each test case by adding prologue and epilogue
instructions. 
The prologue instructions first set the initial execution context, then the instruction stream will be executed. Finally, the epilogue instructions will dump the result for comparison.
We execute the binary on both QEMU and real devices (e.g., RasberryPi 2B). By comparing the execution result, we confirm that
 \code{0xf84f0ddd} is an inconsistent instruction stream.
Specifically, It will generate a \code{SIGILL} signal in a real device while a \code{SIGSEGV} signal in QEMU.

We further analyzed the root cause and successfully disclosed a bug in QEMU. 
According to Figure~\ref{fig:motivation_encoding}, the concrete value of the encoding symbol \code{Rn}
of the instruction stream \code{0xf84f0ddd} is \code{1111}. As shown in the ASL code (line 1) in Figure~\ref{fig:motivation_decoding}.
it is an \code{UNDEFINED} instruction stream.
However, QEMU does not properly check this condition.
Figure~\ref{fig:qemu_decoding} shows the (patched) function (i.e., \code{op\_store\_ri}) in QEMU for decoding the
instruction \code{STR} (immediate).
It continues the decoding process directly from line 12 without any check. 
We then submit this bug to QEMU developers and the patch is issued (as shown in line 8-10).

\begin{figure}[t]
	\centering
	\begin{lstlisting}
    static bool op_store_ri(DisasContext *s, arg_ldst_ri *a, MemOp mop, int mem_idx)
    {
        ISSInfo issinfo = make_issinfo(s, a->rt, a->p, a->w) | ISSIsWrite;
        TCGv_i32 addr, tmp;

        // Rn=1111 is UNDEFINED for Thumb; 
         
    +   if (s->thumb && a->rn == 15) {
    +        return false;
    +   }
    
        addr = op_addr_ri_pre(s, a);

        tmp = load_reg(s, a->rt);
        gen_aa32_st_i32(s, args);
        disas_set_da_iss(s, mop, issinfo);
        tcg_temp_free_i32(tmp);
        op_addr_ri_post(s, a, addr, 0);
        return true;
    }
	\end{lstlisting}

	\caption{\small{Original code of QEMU and the patch for function op\_store\_ri, which aims to translate STR instruction}}

	\label{fig:qemu_decoding}
\end{figure}

\subsection{Test Case Generator}
\label{subsec:case_generator}

In theory, for a 32-bit instruction, there exist $2^{32} = 4,294,967,296$ possible instruction streams,
which are not practical for evaluation. In our work, we need to generate
a small number of representative test cases that cover most behaviors of an instruction.

Specifically, we first parse the encoding schema to retrieve the encoding symbols and then
infer the type for symbols, e.g., a register index or an immediate value. 
After that, we generate an initialized mutation set by pre-defined rules for each type of the symbol
(Table~\ref{tab:initset} shows the detailed rules).
For instance, we generate the maximum, minimum and random values for an immediate value.
Then, we develop a symbolic execution engine to solve the constraints in the ASL code for the decoding and execution
logic. This step can add more values to the mutation set to satisfy
the constraints of the symbols in the ASL program. At last, we remove duplicate values and then generate instruction streams as test cases.

Algorithm~\ref{alg:test_case} shows how we generate the test cases.
For each instruction, ARM provides a XML file to describe the instruction. 
We extract the encoding schemas and the corresponding ASL code for decoding and execution by parsing the XML file.
We first retrieve the encoding symbols ($Symbols$) and constant values ($Constants$) in the encoding schema,
as well as $Constraints$ for the symbolic expression in decoding and execution ASL code (line 2).
We then iterate over the $Symbols$ and generate the $MutationSet$ for each symbol (line 3-4), which will be introduced in detail in Section~\ref{sec:value_coverage}. Note this is the initial mutation set for each symbol. 
For the $Constants$, the $MutationSet$ contains only the fixed value (line 5-6). 
After that, we solve the constraints to generate new mutation set (i.e., $ValueSet$) for each symbol
(line 7-8), which will be introduced in detail in Section~\ref{sec:branch_coverage}.
Then we check whether the solved value for each symbol is in the symbol's $MutationSet$ (line 9).
If not, we append it to the symbols's $MutationSet$ (line 10-11). 
After that, we combine them to get the $MutationSets$ (line 12).

Finally, considering all the possible combinations of the candidates in the $MutationSet$ for each symbol,
we conduct the Cartesian Product on the $MutationSets$ to get the test cases for this specific instruction encoding (line 13).

\begin{algorithm}[t]
	\footnotesize
	\SetKwInput{KwInput}{Input}                
	\SetKwInput{KwOutput}{Output}              
	\DontPrintSemicolon
	\KwInput{The encoding diagram: $I\_Encode$; \\
	The decoding ASL code: $I\_Decode$;\\ 
	The execution ASL code: $I\_Execute$ \\}
	\KwOutput{The generated test cases: $T$;}
	
	\SetKwFunction{FGenerate}{Generate}
	
	\SetKwProg{Fn}{Function}{:}{}
	\Fn{\FGenerate{$I\_Encode$,$I\_Decode$,$I\_Execute$}}{
	    $Symbols$, $Constants$, $Constraints$ = ParseASL($I\_Encode$, $I\_Decode$, $I\_Execute$)\;
	    \For(){$S$ in $Symbols$}{
	         $S.MutationSet$ = InitSet($S$)\;
	        }
	    \For(){$C$ in $Constants$}{
	         $C.MutationSet$ = [ConstantValue]\;
	        }
	    
	    \For(){$C$ in $Constraints$}{
	        ValueSet = SolveConstraint($C$, $Symbols$, $I\_Decode$, $I\_Execute$)\;
	        \For(){$V, S$ in $ValueSet$}{
	        \If(){$V$ not in $S.MutationSet$}{
	        $S.MutationSet$ add $V$\;
	        }
	        }
	    }

        $MutationSets$ = [$S.MutationSet$ +$C.MutationSet$]\;

	    $TestCase$ = CartesianProduct($MutationSets$)\;

		\Return $T$\;
	}

	\caption{The algorithm to generate test cases.}
	\label{alg:test_case}
\end{algorithm}

\subsubsection{Initialize Mutation Set}
\input{tables/initset.tex}
\label{sec:value_coverage}
In the phase of initializing mutation set for each symbol, we consider the types of different symbols and aim
to cover different values for different types of symbols. In particular, we infer the type based on the 
symbol name. For instance, a symbol that represents a register index usually has the name \code{Rd}, 
\code{Rm}, \code{Rn}, etc. As for the immediate value, the symbol name used to be \code{immn} where \code{n}
represents the length of the value. For example, the symbol \code{imm8} represents a 8-bit immediate value.

Table~\ref{tab:initset} shows the rules to initialize the mutation set.
For a register index, we include the \code{PC} register (index 15), \code{R0}, \code{R1}
and random values in the set.
The register \code{R0} and \code{R1} are used to represent the return value for function calls.
As for PC, it can explicitly change the execution flow of the program. Thus, the register index in many instruction encodings
cannot be 15. We include it in the mutation set to cover such cases.
For the immediate value, the maximum and minimum value are the two boundary values that need to be covered.
Apart from this, we randomly select (N-2) values, where N represents the bit length of the symbol.
Note that enumerating all the values for one symbol is not realistic because immediate values have
24 bits, resulting in $2^{24}=16777216$ candidates.

\subsubsection{Solve Constraints}
\label{sec:branch_coverage}

\newcommand{\highlightInListing}[1]{\textit{\textcolor[RGB]{0 139 69}{#1}}}

\begin{figure}
    \centering
    \begin{subfigure}[b]{0.9\linewidth}
        \centering
        \includegraphics[width=\textwidth]{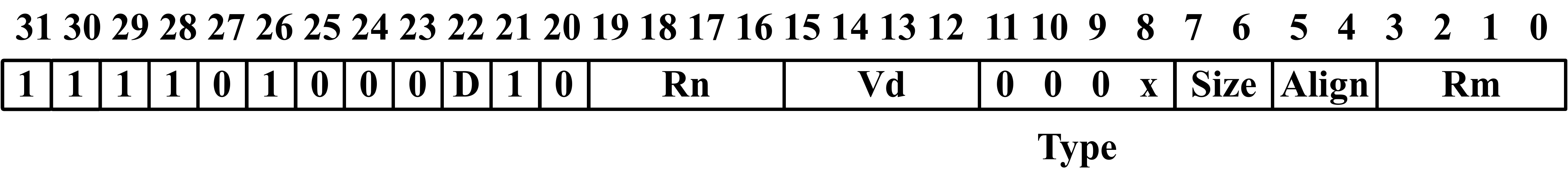}
\caption{Encoding diagram of instruction VLD4 in A32 instruction set}
\label{fig:ASL_encoding}
    \end{subfigure}
\begin{subfigure}[b]{\linewidth}
	\centering
	\begin{lstlisting}
    case type of
        when '0000'
            ~\highlightInListing{inc = 1}~;
        when '0001'
            ~\highlightInListing{inc = 2}~;
    if size == '11' then UNDEFINED;
    alignment = if align == '00' then 1 else 4 << UInt(align);
    ebytes = 1 << UInt(size);  
    elements = 8 DIV ebytes;
    ~\highlightInListing{d = UInt(D:Vd)}~;  
    ~\highlightInListing{d2 = d + inc}~;  
    ~\highlightInListing{d3 = d2 + inc}~;  
    ~\highlightInListing{d4 = d3 + inc}~;  
    n = UInt(Rn);  
    m = UInt(Rm);
    wback = (m != 15);  
    register_index = (m != 15 && m != 13);
    if n == 15 || ~\highlightInListing{d4 > 31}~ then UNPREDICTABLE;
	\end{lstlisting}
	\caption{Decoding code of instruction VLD4 in A32 instruction set}
	\label{fig:ASL_executing}
\end{subfigure}

\caption{\small{Test case generator example.}}

\label{fig:testcase_example}
\end{figure}

Symbolic expressions in ASL code represent the different
execution paths of the instruction.
For instance, \code{d4} in Figure~\ref{fig:testcase_example} is a symbolic expression (\code{$d4 = UInt(D:Vd) + inc + inc + inc$})
that determines whether the instruction is an \code{UNPREDICTABLE} one.
To make our test case representative, the generated test cases should cover as many execution paths as possible.
To this end, we design and implement a symbolic execution engine for the ASL code. 
Specifically, we assign symbolic values for encoding symbols. Then
we generate the symbolic expression for each variable in the ASL code.
After that, we retrieve the constraint of the symbolic expression
and find the concrete values of the encoding symbols that satisfy the constraint and its negation,
e.g., solve the constraints \code{$(d4 > 31) == true$} and \code{$(d4 > 31) == false$}.



Figure~\ref{fig:testcase_example}
shows a concrete example. 
In line 18, there is a symbolic expression \code{d4} and a constraint \code{$d4 > 31$}.
All the related statements (line 3, 5, 10, 11, 12, and 13) are retrieved via backward slicing and
highlighted in the green color.
To solve this constraint, we conduct backward symbolic execution.
Specifically, the symbol \code{d4} is calculated by the expression \code{$d4 = d3 + inc$} in line 13. Thus, the constraint
is converted to \code{$d3 + inc > 31$}. Given the relationship between \code{d3} and \code{d2} in line 11,
and between \code{d2} and \code{d1} in line 11, we further convert it to \code{$UInt(D:Vd) + 3 \times inc >31$}.
The expression \code{UInt(D:Vd)} is converted to \code{$Vd +  2^4 \times D$ } as the symbol \code{Vd} has 4 bits.
Thus, we have the constraint \code{$Vd + 16 \times D + 3 \times inc > 31$}.
Symbol \code{$inc$} is assigned at line 3 and line 5. Thus, the constraint is
\code{$inc == 1\;or\;inc == 2$}. Apart from this, we need to consider the length of each symbol.
Since \code{$D$} is one bit and  \code{$Vd$} has four bits. Their constraints
are 
\code{
$D \geq 0 \; and \; D  < 2$}, 
\code{$Vd \geq 0\;and\;Vd < 16$
}. 

We feed all these constraints to the SMT solver. It returns a solution which is a combination of symbol values that satisfy the constraints.
One possible solution is that \code{$Vd$} is 13, \code{$D$} is 1, and \code{$inc$} is 2.
We then negate the constraint \code{$d4 > 31$} and repeat the above mentioned process.
In this case, the solution is \code{$Vd$} is 0, \code{$D$} is 0 and \code{$inc$} is 1.
Thus, the generated $ValueSet$ contains three symbols and each symbol has two candidate values.
Note \code{$inc$}'s value depends on \code{$Type$}'s value.
As we will also solve the constraint \code{$Type ==$}  `\code{$0000$}' and \code{$Type ==$}  `\code{$0001$}',
the final mutation set of \code{$Type$} must contain the value that can make \code{$inc$} to be
either 1 or 2. Due to the Cartesian Product between each symbol's mutation set, we can always generate the instruction streams
that can satisfy the constraint \code{$d4 > 31$} and its negation.

Note that 
the path explosion in symbolic execution is not an issue for our purpose since the decoding and execution ASL code
has limited constraints, resulting in limited paths.
Meanwhile, we model the utility function calls (e.g., UInt) so that the symbol will not be propagated
into these functions. Our experiment in Section~\ref{sec:rq1} shows that we can generate the test cases within
4 minutes.

\subsubsection{A Demonstration Example}
Table~\ref{tab:testcase_example} describes how we generate all the test cases for instruction \code{VLD4}  in Figure~\ref{fig:testcase_example}. In total, we split the encoding diagram into nine parts including seven
symbols and two constant values (None in the column "Symbol Name").
For constant values, the initialized mutation set has one fixed value.
For other symbols, we initialize the mutation set, which is described in column "Init Mutation Set",
according to  algorithm~\ref{alg:test_case}.
Then we extract the constraints, and find the satisfied values.
Column "Related Constraints" lists the constraints for each symbol. After solving the constraints and their negations, new mutation
sets for each symbol will be generated.
Finally, we have the mutation set for each symbol, which is denoted by column "Final Mutation Set".
We conduct the Cartesian Product between the mutation set of each symbol. In total, we generate $1 \times 2 \times 1 \times 4 \times 6 \times 2 \times 4 \times 3 \times 5 = 5,760 $ test cases for this instruction encoding.

\input{tables/init_mutate.tex}

\subsection{Differential Testing Engine}
\label{subsec:engine}
\subsubsection{Model the CPU}
\label{subsubsec:modelCPU}
The differential testing engine receives the generated test cases (instruction streams), and 
detects inconsistent ones. Formally, given one instruction stream $I$, we denote
the state before the execution of $I$ as the initial state $CPU_I$ and the state after the execution of $I$ as the
final state $CPU_F$.
We denote the CPU $T$'s initial state $CPU_I(T)$ with the tuple $<PC_T, Reg_T, Mem_T, Sta_T>$.  $PC$ denotes the program counter,
which points to the next instruction that will be executed. $Reg$ denotes the registers used by processors while $Mem$ denotes the memory space that the tested instruction $I$ may write into. Note we do not consider the whole memory space due to two reasons.
First, comparing the whole memory space is time- and resource-consuming. Second, memory addresses like stack address are different each time due to specific memory management strategies (e.g., Address Space Layout Randomization). $Sta$ denotes the {status} register, which is $APSR$ in ARM architecture. 
 
We denote the  CPU $T$'s final state $CPU_F(T)$ with the tuple $[PC_T, Reg_T, Mem_T, Sta_T, Sig_T]$. Inside $CPU_F(T)$, all the other attributes except $Sig$ have the same meanings as they are inside $CPU_I(T)$. $Sig$ denotes the signal that the instruction stream $I$ may trigger. If no signal is triggered, the value of $Sig$ is $0$. 

Given the CPU emulator $E$, the real device $R$, our differential testing engine guarantees that  $E$'s initial state $CPU_I(E)$ is equal to the $R$'s initial state $CPU_I(R)$. $CPU_I(E) = CPU_I(R)$ iff:

$$\forall \phi \in <PC, Reg, Mem, Sta>: \phi_E = \phi_R $$

After the execution of $I$, $I$ is treated as an inconsistent instruction stream if the final state $CPU_F(E)$ is not equal to the $R$'s final state $CPU_F(R)$. More formally, $CPU_F(E) \neq CPU_F(R)$ iff:

$$\left\{ 
\begin{array}{ll}
\exists \phi \in [PC, Reg, Mem, Sta]: \phi_E \neq \phi_R & Sig_E = Sig_R = 0 \\
True & Sig_E \neq Sig_R \\
\exists \phi \in [PC, Reg, Sta]: \phi_E \neq \phi_R & Sig_E = Sig_R \neq 0
\end{array}
\right.
$$
 
This is because if the instruction stream $I$ triggers a signal on both CPU emulator $E$ and a real device $R$ with a different signal number,
$CPU_F(E)$ is not equal to $CPU_F(R)$. If the signal number is the same,
we need to compare the $PC$, $Reg$, and $Sta$ as
the instruction stream is not executed normally (otherwise no signal will be raised).
When the $I$ does not trigger a signal  on a CPU emulator $E$ or a real device $R$ ($Sig_E = Sig_R = 0$),
we will compare $Mem$.

Note we do not change any internal logic of the CPU emulator.
Thus, our differential testing engine can be principally applied to different emulators, if necessary.

\subsubsection{Our Strategy}

\begin{figure}
 
	\centering
	\begin{lstlisting}
    int main() {
        register_signals(sig_handler);
        set_init_state();
        __asm__("nop");
        dump_final_state();
        exit();
    }
    
    void sig_handler() {
        dump_final_state();
        exit();
    }
	\end{lstlisting}

\caption{\small{The pseudo code for rendering different test cases.}}

\label{fig:diff_testing}
\end{figure}

To conduct the differential testing, we build a template binary $B$. Given one instruction $I$,
we will generate a new binary $B_I$ by inserting prologue and epilogue instructions. 
Figure~\ref{fig:diff_testing} shows  pseudo code. We first register the signal handlers (line 2)
to capture different signals.
This is because the test instruction may trigger exceptions such as illegal encoding, memory error, etc.

To make the initial state consistent (line 3), we set the value of general purpose registers to zero
except the  frame register (R11), the stack register (R13), the  link register (R14),  and the PC. This is because these registers have specific functionalities. For instance,
link register is used to save the return address (a non-zero value) while PC can influence the execution flow.

After setting up the initial state, a test case (an instruction stream) will be executed.
We set the instruction \code{nop} with inline assembly (line 4). Given one test instruction stream,
we statically rewrite the binary and change the \code{nop} instruction to the test instruction stream to generate a new binary.
After the execution, we need to dump the CPU state (line 5 and line 10) so that we can compare the execution result. 
Note the test instruction stream could trigger a signal, thus, we also need to dump the result in signal handlers.
 
For register values, we push them on the stack and then write them into a file.
For the memory, we check the test instruction with Capstone~\cite{capstone}, analyze the instruction to see whether it will
write a value into a memory location. If so, we load the memory address, and push it on the stack for later inspection.
Finally, we compare the result collected from the emulator and a real device. If the instruction stream results in a different
CPU final state, ($CPU_F(E) \neq CPU_F(R)$), it will be treated as an inconsistent instruction stream.

%% file: tables/initset.tex
\begin{table}[]
\centering
\footnotesize
\caption{\small{The rules of initializing the mutation set.}}

\begin{tabular}{cc}
\toprule
Type of Symbol Name             & Mutation Set                                                                                                                                       \\ \hline \hline
Register Index                          & 0 (R0); 1 (R1); 15 (PC); Random index values                                                                                                                                              \\ \hline
Immediate Value in N bits          & \begin{tabular}[c]{@{}l@{}}Maximum value: 2\textasciicircum{}N -1; Minimum value: 0; \\ (N-2) Random Value from the enumerated values\end{tabular} \\ \hline
Condition                          & "1110" (Always execute)                                                                                                                            \\ \hline
Others in 1 bit                    & "0"; "1"                                                                                                                                           \\ \hline
Others in N bit (N \textgreater 1) & N random value from the enumerated values                                                                                                          \\ \bottomrule
\end{tabular}
\label{tab:initset}
\end{table}

%% file: tables/init_mutate.tex
\begin{table*}[htb!]
\caption{ \small{The generated mutation set for each symbol of instruction VLD4 in Figure~\ref{fig:testcase_example}}}

\centering
\footnotesize
\scalebox{0.74}{
\begin{tabular}{cccccccccc}
\toprule
\begin{tabular}[c]{@{}c@{}}Symbol \\ Name\end{tabular} & \begin{tabular}[c]{@{}c@{}}Bit \\ Length\end{tabular} & \begin{tabular}[c]{@{}c@{}}Start \\ Offset\end{tabular} & \begin{tabular}[c]{@{}c@{}}End \\ Offset\end{tabular}  & Type             & Init Mutation Set & Related Constraint & \begin{tabular}[c]{@{}c@{}}Set Added by Solving \\ Constraints and Their Negations \end{tabular}    & Final Mutation Set & Set Size \\ \hline
\hline
None & 9 &23 & 31 & Fixed Value & "111101000"&NA&NA&"111101000" &1\\ \hline
D           & 1   &22&22       & Others in 1 bit  & "0", "1" & $d4 > 31$ & "0", "1"  &   "0", "1" &2 \\ \hline
None & 2 & 20 & 21 & Fixed Value &  "10"  & NA& NA&"10"&1\\\hline
Rn          & 4    &  16&19    & Register Index         & \begin{tabular}[c]{@{}c@{}}"0000","0001", \\ "0110","1111" \end{tabular}
& n == 15 & "0000","1111" &\begin{tabular}[c]{@{}c@{}}"0000","0001", \\ "0110","1111"\end{tabular} &4\\ \hline
Vd          & 4   &12&15       & Others in 4 bit  & \begin{tabular}[c]{@{}c@{}}"0101","0110", \\ "1001","1100"\end{tabular}
& $d4 > 31$ &  "0000","1101"   & \begin{tabular}[c]{@{}c@{}}"0000","0101","0110", \\ "1001","1100","1101"\end{tabular} &6\\ \hline
Type        & 4    & 8 & 11      & Others in 4 bit  &  "0000","0001"   & \begin{tabular}[c]{@{}c@{}}Type == '0000' \\ Type == '0001'\end{tabular} & "0000","0001"   &   "0000","0001" &2  \\\hline
Size        & 2    & 6&7      & Others  in 2 bit &  "01","10"  & Size == '11'  &  "00","11"    & "00","01","10","11"  &4 \\ \hline
Align       & 2   &4&5       & Others in 2 bit  &  "00","11"   & Align == '00' &   "00","01"    & "00","01","11" &3 \\ \hline
Rm          & 4   &0&3       & Register Index        &\begin{tabular}[c]{@{}c@{}} "0000","0001",\\"0111","1111"\end{tabular}    & \begin{tabular}[c]{@{}c@{}}m != 15\\ m != 13\end{tabular} & "0000","1101","1111"  &  \begin{tabular}[c]{@{}c@{}} "0000","0001",\\"0111","1101","1111" \end{tabular} &5  \\

\bottomrule

\end{tabular}}
\label{tab:testcase_example}
\end{table*}

%% file: 04_implementation.tex
\subsection{Implementation Details}
We have implemented a prototype system \sysname using Python, C and ARM assembly.
In particular, we implement the  test case generator in Python. We parse the ASL code,
extract the lexical and syntactic information with regular expressions. We use Z3~\cite{z3} as the
SMT solver to solve the constraints.
The differential testing engine is implemented in C and assembly code with some glue scripts in Python.
Specifically, the initial state setup and the execution result dumping is implemented with inline assembly.
In total, \sysname contains $5,074$ lines of Python code, $220$ lines of C code, and $200$ lines of assembly code.

%% file: 05_evaluation.tex
\section{Evaluation}

In this section, we evaluate \sysname by answering the following three research questions.
\begin{itemize}[leftmargin=*]
 \setlength{\itemsep}{1pt}
	\item \textbf{RQ1:} Is \sysname able to generate sufficient test cases?
	\item \textbf{RQ2:} Is \sysname able to detect inconsistent instructions? What are the root causes of these inconsistent instructions?
	\item \textbf{RQ3:} What are the possible usage scenarios of inconsistent instructions?

\end{itemize}

\subsection{Sufficiency of Test Case Generator \textbf{(RQ1)}}
\label{sec:rq1}
\input{tables/eva_testcases.tex}
We generate the test cases according to ARMv8-A manual, which introduces ASL. Specifically, the manual includes four different instruction sets. In AArch 64 mode, A64 instruction set is supported.
For the AArch 32 mode, it consists of three different instruction sets. They are ARM32 with 32-bit instruction length (A32),
Thumb-2 with instruction length of mixed 16-bits and 32-bits (T32), and Thumb-1 with 16-bit instruction length (T16). They are also supported by previous ARM architectures (e.g., ARMv6, ARMv7). To locate the inconsistent instructions in different ARM architectures, we generate the test cases for all the instruction sets. 

The generated test case is sufficient. Table~\ref{tab:testcase} shows the statistics of the generated instructions. For A64, we generate around 1 million instructions, which cover all the $839$ instruction encodings in $581$ instructions. In the decoding and executing ASL code, we solved $3,436$ constraints that are related to encoding symbols.

We noticed that both A32 and T32 have around 5 hundred instruction encodings and more than 800 thousands instructions are generated.
For T16, the generated instruction streams are rather small due to the small number of instruction encoding schemes and limited
instruction length. 
Overall, all the generated instruction streams are valid instruction streams, which means they meet the encoding schema of one instruction encoding. Meanwhile, all the instruction encodings and instructions are covered. Furthermore, more than 12 thousand constraints, which are related to encoding symbols, are solved, indicating the multiple behaviors of the instructions are explored. 

To demonstrate the effectiveness of the test case generation algorithm, we randomly generate some instruction streams. To make the comparison fair, we generate the same number of test cases for each instruction set. 
Table~\ref{tab:random_testcase} shows the result. We repeat the randomly generated process for 10 times. Then we check whether the generated instructions are valid instruction streams or not. If they are the valid instruction streams, we calculate how many instruction encodings, how many instructions, and  how many constraints are covered by these instruction streams. We noticed that only 37.3\% generated instruction streams are valid instruction streams, which means all the other instructions are illegal instructions and they are not effective to test the potential different behaviors between real devices and CPU emulators. Among the valid instruction streams, it can only cover 54.5\% instruction encodings and 51.4\% instructions.
Nearly a half of instructions can not be covered with the randomly generated instruction streams. Specifically, many of the T32 instructions cannot be covered with randomly generated instructions, which means many of these instructions have fixed value. 
As for the coverage of constraints, 62.6\% constraints are covered while the left 37.4\% constraints can not be explored, resulting in a relatively low behavior space. 

\input{tables/random_testcases_new.tex}

\begin{tcolorbox}[size=title]
{\textbf{Answer to RQ1: } \sysname can generate sufficient test cases, which  are all valid instruction streams and can cover all instruction encodings and instructions. On the contrary, Only 37.3\% of the same number of randomly generated instruction streams are valid instruction streams. Furthermore,  45.5\% instruction encodings, 48.6\% instructions, and 37.4\% constraints cannot be explored by these randomly generated instructions.}
\end{tcolorbox}

\subsection{Differential Testing Results and Root Causes \textbf{(RQ2)}}
\label{sec:rq2}
\input{tables/testing_result.tex}
With the generated test cases in four different instruction set. We feed them into our differential testing engine to locate the inconsistent instructions. Table~\ref{tab:testing_result} shows the result.

\smallskip
\noindent \textbf{Experiment Setup}\tab
In total, we conduct the differential testing between QEMU (version 5.1.0) and four different ARM architecture versions (i.e., ARMv5, ARMv6, ARMv7-a, ARMv8-a). For each ARM architecture version, we select one real device.  Specifically, we select devices OLinuXino iMX233 for ARMv5, RaspberryPi Zero for ARMv6, RaspberryPi 2B for ARMv7-a, and Hikey 970 for ARMv8-a. To make the differential testing fair, we use the same or similar CPU model between the emulator and real device. Note the CPU model for Hikey 970 is A73 while the most advanced CPU model supported by QEMU is A72, which is selected. However, they two both share the same instruction set (i.e., ARMv8-a).
For ARMv5, only A32 instruction set is supported.  Since QEMU does not support Thumb2 for ARM1176 of ARMv6, we only test the A32 instruction set on ARMv6. For ARMv7, all the instruction set is supported except A64. Since the T16 instruction has a rather small number of set. We combine the T16 and T32 in the testing process. For ARMv8-A, only A64 instruction set is supported in user-level programs. 
The "Generated Instruction Streams",  "All Instruction Encodings", and "All Instructions " in Table~\ref{tab:testing_result}  are from the "GIS", "AE", and "AI" in Table~\ref{tab:testcase}, respectively. In total, it takes around 2700 seconds of CPU time for QEMU, which is run on the Intel i7-9700 CPU. For the real devices, the CPU time cost ranges from 5276 seconds to 46238 seconds (around 13 hours), depending on the specific devices. Thanks to the representative test cases, the differential testing for all the test cases can be finished within acceptable time.

\smallskip
\noindent \textbf{Testing Result} \tab
Among all the test cases, some of them may read from or write into SP and FP registers. SP and FP are used to store function parameters and local variables. They are influenced by the memory management strategy (e.g., Address Space Layout Randomization) of the whole system and are different for each run.  Meanwhile, some instruction streams may change these two registers and result in crash of the whole test binary. Thus, instruction streams that read from or write into these two registers are filtered. Apart from this, some instruction streams are branch instructions. These instructions 
may execute normally on both emulator and real devices. Then inconsistent behaviors may occur due to the other instructions as the branch instructions change the execution flow of the binary. We also filter these instructions if they execute normally. For the left instruction streams, we noticed around 90\% instruction streams (from 88.9\% to 94.3\% for each architecture version)  are consistent for four different ARM architectures. However, 
there are still thousands of  inconsistent instruction streams. In particular, 155,642 inconsistent instruction streams are found, owning to 5.6\% of the whole test cases.
Note one instruction may be tested in different architectures (e.g.,A32 instruction set in ARMv5, ARMv6, and ARMv7), the number in column "Overall" is the union of the other columns, which means the instruction stream can cause inconsistent behavior in at least one architecture.
Furthermore, these inconsistent instruction streams cover 600 different instruction encodings and 511 instructions, owning 30\% and 47.8\% of the all instruction encodings and instructions, respectively.

\smallskip
\noindent \textbf{Inconsistent Behavior}\tab
We further analyze the inconsistent instruction streams and categorize them according to our modeled CPU. We noticed nearly half of the inconsistent instruction streams (i.e., 47.3\%) raise different signal numbers during the execution. Meanwhile, 42.1\% inconsistent instruction streams would trigger signals in real devices while they can be executed normally in QEMU. The percentage of inconsistent instruction streams drops to 4.6\% if they can trigger signals on QEMU but cannot on real devices, which demonstrate that QEMU is more tolerant compared with real devices for many instructions. Overall, the above mentioned three cases ($Sig\_E \ne Sig\_R$) take 94.0\% (47.3 + 4.6 + 42.1) of all the inconsistent instruction streams. A small number of instruction streams  may (i.e., 3.7\%) or may not (i.e., 1.5\%) trigger the same signals but have different register or memory values. They mainly because of the UNPREDICTABLE conditions.  The left 2.8\% inconsistent instruction streams are due to the other problems. These instruction can make the emulator or real devices crash or stuck. 

Note instruction streams generated from one instruction encoding can result in different inconsistent behaviors due to the detail decoding and executing logic. According to our experiments, the inconsistent behaviors with more inconsistent instruction streams usually cover more instruction encodings and instructions. For example, $458$ different instruction encodings from $401$ instructions are covered by the $73,619$ inconsistent instruction streams that trigger different signals.

\begin{figure}
 
	\centering
	\begin{lstlisting}
    boolean AArch32.ExclusiveMonitorsPass(bits(32) address, integer size)
    // It is IMPLEMENTATION DEFINED whether the  
    // detection of memory aborts happens before or   
    // after the check on the local Exclusive Monitor.  
    // As a result, a failure of the local monitor can  
    // occur on some implementations  even if the
    // memory access would give an memory abort.
        ...
        return 
	\end{lstlisting}
\vspace{-1em}
\caption{\small{Two different implementations are defined in the annotation of function ExclusiveMonitorsPass, which is called by many instructions' executing code}}
\label{fig:exclusive}
\end{figure}

\smallskip
\noindent \textbf{Root Cause}\tab Based on the above mentioned inconsistent behavior, we  explore the root cause of the inconsistent instructions. First, there are implementation bugs of QEMU. We discovered 4 bugs in total, which come from 586 inconsistent instruction streams and 13 instruction encodings. Some of the bugs are related to very common instructions. For example, many load and store instructions (e.g., \code{LDRD}, \code{STRD}, \code{LDM}, \code{STM}, etc.) in A32 instruction set should check the alignments mandatory while QEMU does not check. Instructions like \code{STR}, \code{BLX} may be undefined instructions in specific cases, which should raise SIGILL signal, even they may meet the corresponding encoding schema. However, QEMU does not follow the specification. We also noticed one instruction (i.e., \code{WFI}) that can make QEMU crash. \code{WFI} denotes waiting for interrupt and is usually used in system-mode emulation. However, ARM manual specifies that it can also be used in user-space. QEMU does not handle this instruction well and an abort will be generated during user-space emulation. After our report, all of these bugs are confirmed by developers and are in patching process. This also demonstrates the capability of \sysname in discovering the bugs of the emulator implementation.

Apart from the bugs, many other inconsistent instruction streams are due to the undefined implementation in the ARM manual. There are three different kinds of undefined implementation. The first one is UNPREDICTABLE, which is introduced in Section~\ref{subsec:motivation_example}. UNPREDICTABLE leaves open implementation decision for emulators and processors, which can result in inconsistent instructions. UNPREDICTABLE is the major reason  and it takes 88.6\% for all the inconsistent instruction streams. The second is Constraint UNPREDICTABLE ("Cons\_UNPRE" in Table~\ref{tab:testing_result}). Constraint UNPREDICTABLE provides candidate implementation strategies and the developer or vendor can choose from one of them, which only exists in A64 instructions. 
The last one is defined in the annotation part of the ASL code ("Annotation\_Def" in Table~\ref{tab:testing_result}). Figure~\ref{fig:exclusive} shows an example. In the function \code{ExclusiveMonitorsPass}, which is called by the executing code of  instruction \code{STREXH}, there is an annotation for the implementation. Note the check on the \textit{local Exclusive Monitor} would update the value of a register. Thus, if the detection of memory aborts happens before the check, the value of the register would not be updated while the detection happens after the check can update the value, resulting in different register value. 

\input{tables/android.tex}

Some specific instructions (i.e., \code{BKPT}) can trigger SIGTRAP signal in QEMU while make the real devices stuck. As we cannot verify the detail implementation logic of real devices without the design specification, we left it to be others.

\begin{tcolorbox}[size=title]
{\textbf{Answer to RQ2: } \sysname can detect inconsistent instructions. In total, 155,642 inconsistent instruction streams are found, which covers  30\% (i.e.,600/1998) instruction encodings and 47.8\% instructions (i.e., 511/1070). The implementation bugs of QEMU and the undefined implementation in ARM manual are the major root causes. Four bugs are discovered and confirmed by QEMU developers. These bugs influence 13 instruction encodings including commonly used instructions (e.g., \code{STR}, \code{BLX}).
}
\end{tcolorbox}

\subsection{Applications of Inconsistent Instructions \textbf{(RQ3)}}
According to the evaluation result in Section~\ref{sec:rq2}, there are many inconsistent instructions (i.e., 47.8\%) in ARM architectures. These instructions can be used to detect the existence of emulators. Furthermore, detecting emulator can prevent the binary from being analyzed or fuzzed, which is known as anti-emulation and anti-fuzzing technique.

\subsubsection{Emulator Detection}

\begin{figure}
 
	\centering
	\begin{lstlisting}
	  void sig_handler(int signum) {
        record_execution_result(i++);
        siglongjmp(sig_env, i);
    }

    Bool JNI_Function_Is_In_Emulator() {
        register_signals(sig_handler);
        i = sigsetjmp(sig_env,0);
        switch (i){
            case 1:
                execute(inconsistent_instruction_n);
                record_execution_result(i++);
                longjmp(sig_env,i++);
            case 2:
            ...
            case n:
        }
        return compare_result();
    }
	\end{lstlisting}
\vspace{-1em}
\caption{\small{Pseudo code of the native code for detecting the emulator.}}

\label{fig:emulator_detect}
\end{figure}

The inconsistent instructions can be used to detect emulators. We use the inconsistent instructions  for emulator detection. Considering the popularity of Android systems, we target Android applications. 
Specifically, we build a native library by using the inconsistent instructions. For each encoding of one inconsistent behavior, we select one instruction stream.

Figure~\ref{fig:emulator_detect} shows the pseudo code of the library. Function \textit{JNI\_Func\-tion\_Is\_In\_Emulator} (line 6) returns True if the emulator is detected.
Inside the function, we register signal handlers for different signals (line 7). After the execution of each instruction stream, we will record the execution result either in the signal handler (line 2) or after the execution (line 12), which depends on the execution result. Then we use the function \textit{longjmp} (line 13) or \textit{siglongjmp} (line 3) to jump back to the place where calling \textit{sigsetjmp} (line 8). Note the return value of \textit{sigsetjmp} depends on the parameter of the \textit{longjmp} or \textit{siglongjmp}. Thus, the \textit{switch} would check the value of \textit{i}, which is the return value of \textit{sigsetjmp}, repeatedly. As \textit{i} would increase one after the execution of one instruction stream, we can execute hundreds of instruction streams in one function by adding corresponding \textit{case} conditions. Each instruction stream can make an equal contribution to the final decision on whether the current execution environment is in real devices or emulators. Finally, if more instruction streams decide the application are running inside an emulator, the \textit{compare\_result()} will return True and vice versa.

We automatically generate the test library with template code and build three Android apps for different instruction set (one for A64, one for A32, and one for T32 \& T16).
 We run the applications on 12 different mobiles from 6 different vendors. These mobiles also use different CPUs, which decides how an instruction stream should be executed. Meanwhile, we run the applications in the Android emulator provided by Android studio (Version 4.1.2). We compare the result running in each real mobiles with the one in emulators. If the function \textit{JNI\_Function\_Is\_In\_Emulator} returns True in emulator and returns False in real mobiles. We consider it will successfully detect the emulator. Table~\ref{tab:android} shows the evaluation result,
by testing the three Android apps (one for A64, one for A32, and one for T32 \& T16)  in 12 mobiles, all the mobile apps can detect the existence of emulator and real mobiles successfully.

\subsubsection{Anti-Emulation}
\label{sec:anti_emu}

\begin{figure}[t]
	\centering
	\includegraphics[width=0.3\linewidth]{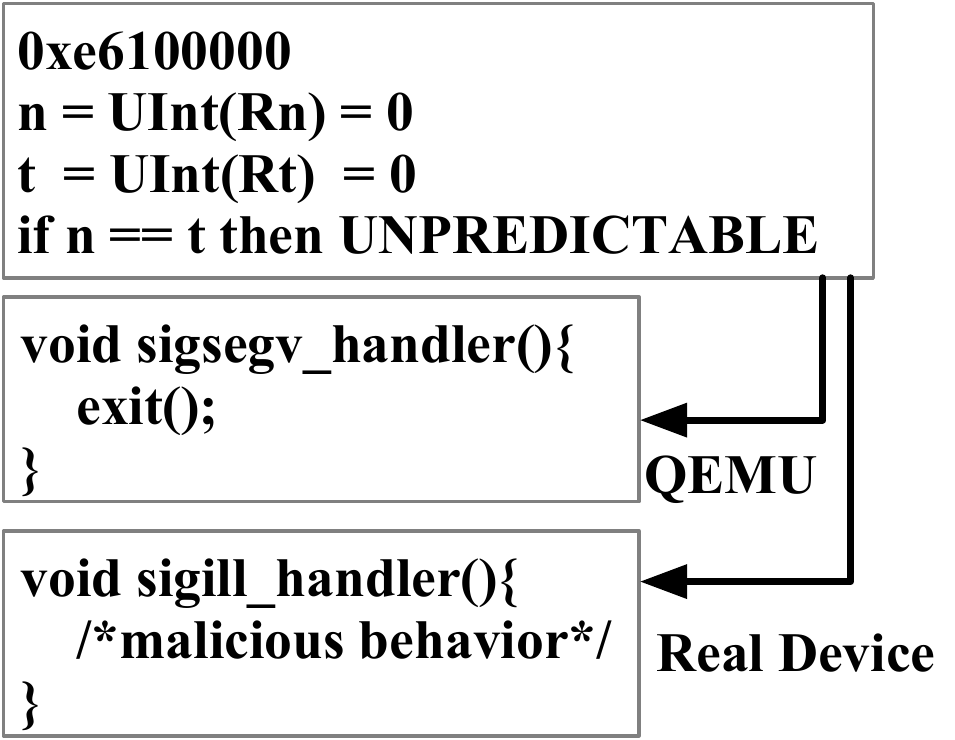}

	\caption{\small{Inconsistent instruction can prevent the malicious behavior being detected by emulators}}

	\label{fig:panda}
\end{figure}
Anti-emulation technique is important.
On the attacker's side, it can be proposed to increase the bar for analyzing the malware so that the defense mechanism can be developed slower. On the defender's side, commercial software needs to protect the core functionality and algorithms from being analyzed. Thus, it is widely used in the wild~\cite{chen2008}.

The inconsistent instructions can be used to conduct anti-emula\-tion and can prevent the malware's malicious behavior being analyzed.  We demonstrate how the inconsistent instruction can be used to hide the malicious behavior.

We use one of the state-of-the-art dynamic analysis platforms (i.e., PANDA~\cite{panda}) to demonstrate the usage. PANDA is built upon  QEMU and supports taint analysis, record and replay, operating system introspection, and so on. We port one of the open source rootkits (i.e., Suterusu~\cite{suterusu}) to Debian 7.3. We register two different signal handlers for SIGILL and SIGSEGV, respectively. Then we instrument one instruction stream (i.e., 0xe61000000). This is a LDR instruction encoding in ARM instruction set. According to the encoding schema, n equals to t and both these two symbols' values are zero. The  ASL code of decoding would check whether n equals to t. If so, it should be the UNPREDICTABLE behavior. Real devices think this is an illegal instruction stream and will raise the SIGILL signal while QEMU tries to execute the instruction stream. Then SIGSEGV will be raised as the address pointed by R0 cannot be accessed. In this case, the malicious behavior will only be triggered in real devices. Meanwhile, when we use the PANDA to analyze the malware, no malicious behavior will be monitored and the program will exit inside the \textit{sigsegv\_handler}.

\subsubsection{Anti-Fuzz}
\label{sec:anti_fuzz} 
\begin{figure}
 
	\centering
	\begin{lstlisting}
	    0x10000: e51b3008 LDR r3,[fp,#-8]
      0x10004: e1a03000 MOV r3,r0
      0x10008: e7cf0e9f BFC r0, #0xf, #1
      // BFC instruction is to clear specific bits
      // e7cf0e9f is an UNPREDICTABLE encoding
      // e7cf0e9f is executed normally in real device 
      // e7cf0e9f triggers SIGILL signal on QEMU
      0x1000c: e1a00003 MOV r0,r3 
      0x10010: e50b3008 STR r3,[fp,#-8]
    
	\end{lstlisting}
\vspace{-1em}

\caption{\small{Instrumented instruction streams for anti-fuzzing.}}
\label{fig:fuzz_snip}
\end{figure}

Fuzzing is widely used to explore the zero-day vulnerabilities. To help the released binaries from being fuzzed by attackers, researchers utilize anti-fuzzing techniques~\cite{jung2019fuzzification,guler2019antifuzz}. 
Considering that many new binary fuzzing frameworks are based on QEMU, the inconsistent instructions can be used by developers as a mitigation approach towards fuzzing technique.

We demonstrate how the inconsistent instructions can be used to conduct anti-fuzzing tasks with a relatively low overhead and high decreased coverage ratio.
 
Figure~\ref{fig:fuzz_snip} shows a snippet of assembly code instrumented into the release binary. In address 0x10008, the instruction \code{BFC} is used to clear bits for register R0.  Note we move the value of R0 to R3 before the instruction \code{BFC} and return it back after the execution of \code{BFC}. This can guarantee the instrumented instructions will not affect the execution of the binary on the real device. The instruction stream  0xe7cf0e9f results in an  UNPREDICTABLE condition. It can be executed normally in real devices while triggering a signal on QEMU.

We developed a GCC plugin to instrument the above mentioned inconsistent instruction streams at each function entry and apply this plugin on three popular used libraries (i.e., libtiff, libpng, and libjpeg) during the compilation process to generate released binaries.

Table~\ref{tab:antifuzz-overhead} shows the space and runtime overhead of the instrumented binary compared with the normal (non-instru\-mented) one. 
The space overhead is measured by comparing the binary size.
For runtime overhead, we measure it by running test suites on both binaries and comparing the cost of time.
We noticed that the instrumented binary  imposes negligible space and runtime overhead to the binary.
The average space overhead for the protected binary is around 4\%, and the runtime overhead is less than 1\%.

\begin{table}[]
\footnotesize
\centering
\scalebox{0.9}{
\begin{threeparttable}
\caption{\small{Overhead information of anti-fuzzing.}}
\label{tab:antifuzz-overhead}

\begin{tabular}{cccc}
\toprule
Library \tnote{1}
& Test Suite\tnote{2} & Space Overhead & Runtime Overhead \\ \hline\hline
libpng (readpng)  & built-in (254)   & 4.0\% (+7KB)       & 0.52\%                        \\
libjpeg (djpeg) & GIT~\tnote{3}~~(97)        & 4.3\% (+8KB)      & 0.61\%                        \\
libtiff (tiffinfo) & built-in (61)   & 2.2\% (+8KB)       & 0.59\%                        \\ \hline
Overall &            & 3.5\%             & 0.57\%   \\
\bottomrule
\end{tabular}

\begin{tablenotes}
\footnotesize
\item[1] {All libraries are compiled using default compile parameters.}
\item[2] {The test inputs for libjpeg is taken from Google Image Test Suite.}
\item[3] {The number of test inputs in test suite is shown in the bracket.}
\end{tablenotes}

\end{threeparttable}
}
\end{table}

\begin{figure*}[htp]
\centering
\begin{subfigure}[h]{0.28\textwidth}
    \centering
    \includegraphics[width=\textwidth]{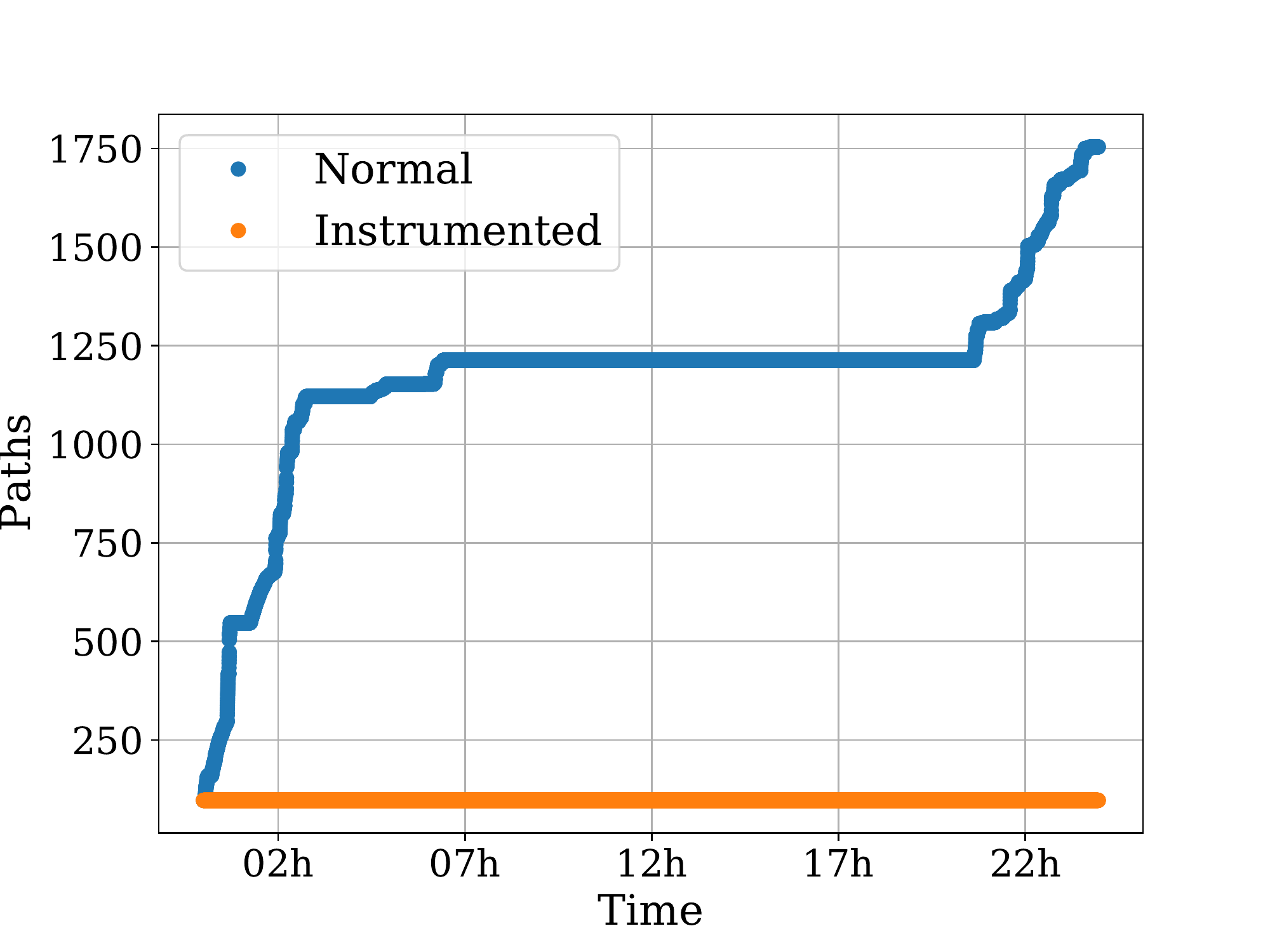}
    \caption{libjpeg}
\end{subfigure}
\begin{subfigure}[h]{0.28\linewidth}
	\centering
	\includegraphics[width=\textwidth]{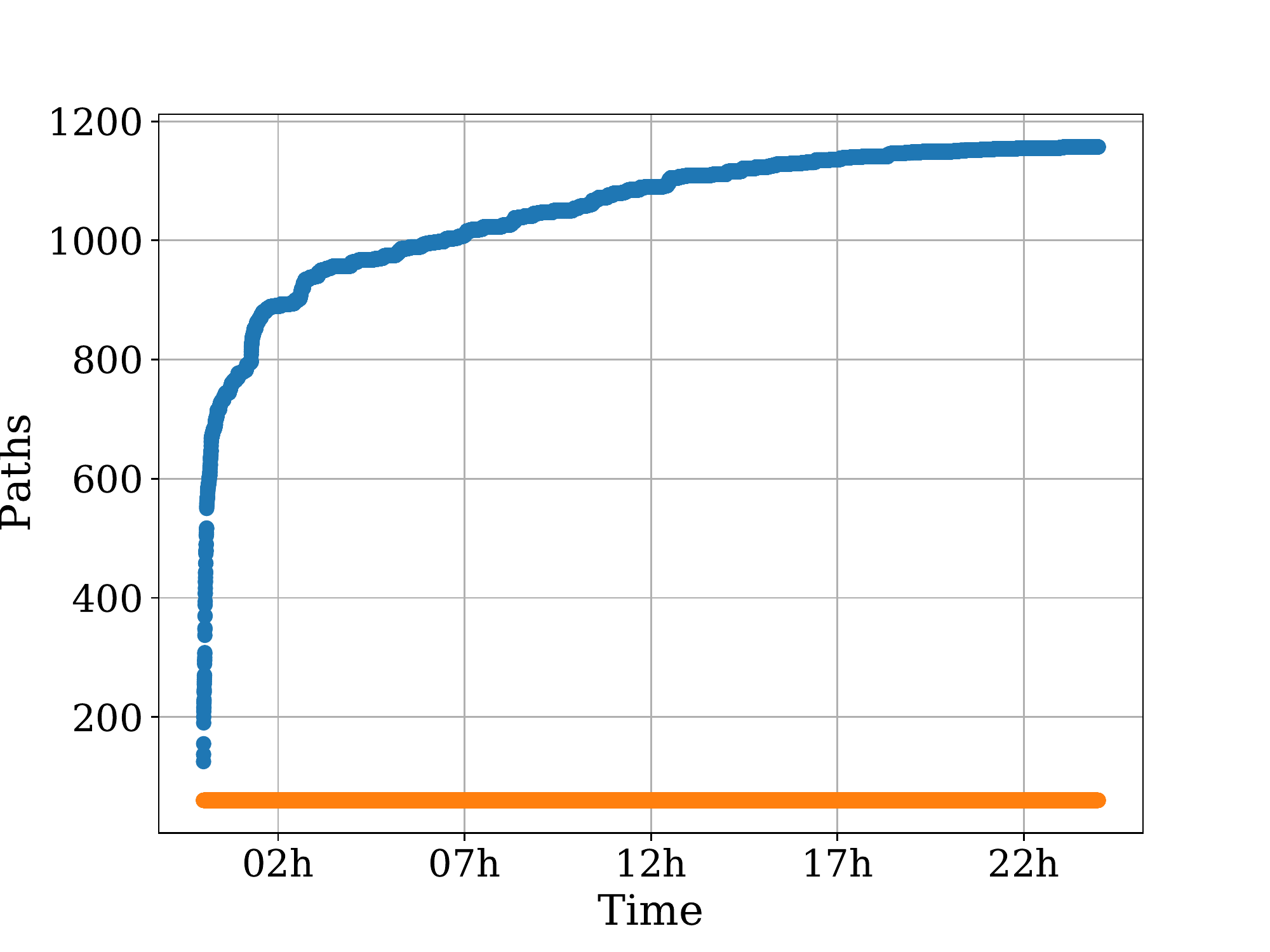}
    \caption{libpng}
\end{subfigure}
\begin{subfigure}[h]{0.28\linewidth}
	\centering
	\includegraphics[width=\textwidth]{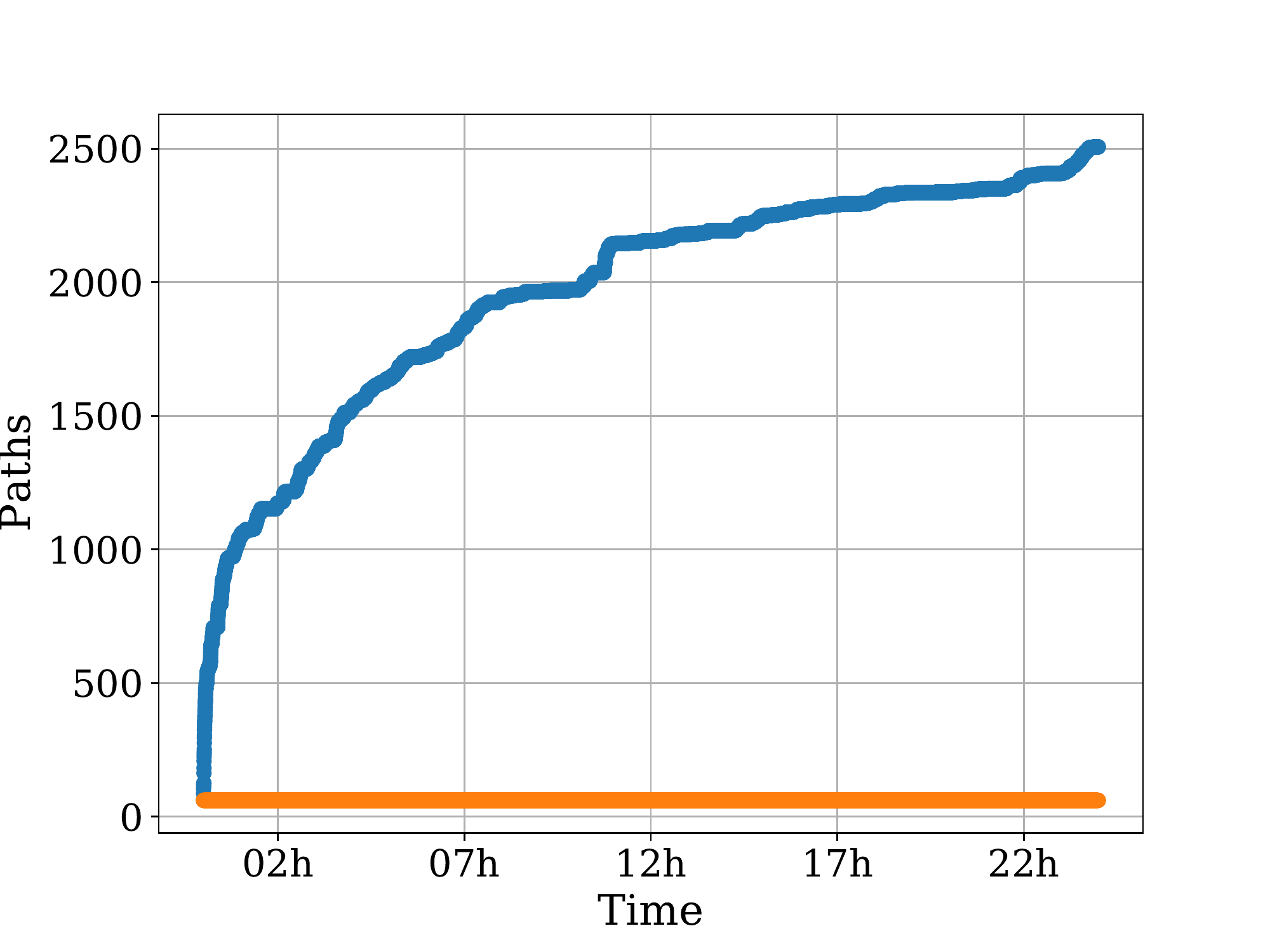}
    \caption{libtiff}
\end{subfigure}

\caption{The result of Anti-Fuzzing experiment on three libraries. The blue lines show the coverage over 24 hours of fuzzing. 
The orange line shows the coverage for instrumented binaries, which decreases due to failed executions of QEMU. }

\label{fig:antifuzz-coverage}
\end{figure*}

We then measure the functionality of anti-fuzzing. We fuzz the instrumented binaries and the normal ones with AFL-QEMU (version 2.56b) for 24 hours. 
The seed corpus is the test suite used for each library in Table~\ref{tab:antifuzz-overhead}. We collect the coverage information for the instrumented and the normal ones. Figure~\ref{fig:antifuzz-coverage} shows the results. 
It is easy to see that the coverage for instrumented binaries cannot increase (because QEMU fails to execute binaries correctly), while the normal ones will increase with the fuzzing time.

Note this is to demonstrate the ability of inconsistent instructions on anti-fuzzing tasks. How to stealthily use these instructions is out of our scope. It is not easy for attackers to precisely recognize all the inconsistent instructions, which will be discussed in detail (Section~\ref{sec:discussion}).

\begin{tcolorbox}[size=title]
{\textbf{Answer to RQ3: } The inconsistent instructions are useful. We demonstrate that the inconsistent instructions can be used to detect the existence of the CPU emulator and  prevent the malicious behavior from being monitored by dynamic analysis frameworks. Furthermore, the path coverage of programs fuzzed in emulators can be highly decreased with the help of inconsistent instructions.
}
\end{tcolorbox}

%% file: tables/eva_testcases.tex
\begin{table*}[]
\centering
\footnotesize
\caption{\small{The statistics of the generated instructions. "GIS" denotes the number of generated instruction streams. "VIS" denotes the number of valid instruction streams, which means they match the instruction encodings. "VISR" denotes the percentage of dividing "VIS" by "GIS". "AE" denotes the number of all instruction encodings. "CE" denotes the number of covered instruction encodings by the generated instruction streams. "CER" denotes the percentage of dividing "CE" by "AE". "AI" denotes the number of all instructions. "CI" denotes the number of covered instruction encodings by the generated instruction streams. CIR denotes the percentage of dividing "CI" by "AI". Note one instruction may have different instruction encodings for different instruction set. The total instruction for instruction set A32, T32, and T16 is 489.}}

\begin{tabular}{cc|ccc|ccc|ccc|c}
\toprule
\begin{tabular}[c]{@{}c@{}} Instruction \\ Set\end{tabular}  &Time (s)  &   GIS&  VIS& VISR & AE & CE &CER &AI & CI & CIR &\begin{tabular}[c]{@{}c@{}}Solved  \\ Constraints \end{tabular}  \\ 
\hline
\hline
A64    & 70.51     & 1,094,700   & 1,094,700 & 100                                                             & 839    & 839 & 100                                                          &581 &581 &100    &   3,436                                                        \\ \hline
A32   &75.05   & 870,221       & 870,221    &100                                                    & 550   & 550   & 100                                                           &  481 &  481 &  100   & 4,718                                                           \\ 
T32 &74.58& 808,770        & 808,770   &100                                                     & 531   & 531       & 100                                                          & 451  &451& 100 & 4,425                                                      \\ 
T16 &2.32& 958      & 958        &100                                                     & 78     & 78     & 100                                                       & 68&68&100 & 122                                                       \\ \hline
Overall& 222.46 &2,774,649   &2,774,649& 100 & 1,998 & 1,998 & 100 & 1,070 &1,070&100& 12,701\\ 

\bottomrule
\end{tabular}
\label{tab:testcase}
\end{table*}

%% file: tables/random_testcases_new.tex
\begin{table}[]
\centering
\footnotesize
\caption{\small{The statistics of the generated instructions in random. "IS" denotes instruction set. "Ave" denotes average value. For each instruction set and the overall result, "R" inside column "VIS", "CE", "CI", and "CC" denotes the percentage of dividing "Ave" inside the same column by "VIS", "CE", "CI", and "Solved Constraints" in Table~\ref{tab:testcase}, respectively. "GIS", “VIS”, "CE", "CI" denotes the same meaning illustrated in Table~\ref{tab:testcase}. "CC" denotes covered constraints.}
}

\begin{tabular}{cc|cc|cc|cc|cc}
\toprule
IS  & GIS & \multicolumn{2}{c|}{VIS} & \multicolumn{2}{c|}{CE} & \multicolumn{2}{c|}{CI} & \multicolumn{2}{c}{CC} \\ \cline{3-10} 
                           &     &Ave    &
                   R &   Ave    &
                   R &  
                    Ave &    
                   R &  
                   Ave &    
                   R 
                   \\ \hline
                    \hline
A64             &1,094,700                & 421,645  & 38.5    & 265           &31.6 & 178& 30.6        & 934    & 27.2     \\ \hline
A32            &870,221                      & 578,845  & 66.5              & 415   & 75.5    & 361& 75.1         & 3,725   &  79.0   \\ 
T32          &808,770                      & 34,598  & 4.2            & 351   &  66.1  & 283 & 62.7      & 3,203   &  72.3    \\
T16            &958                  & 796         &   83.0             & 57  & 73.1    & 49&  72.1           & 84     &   68.9   \\ \hline 
Overall & 2,774,649  & 1,035,884 & 37.3 &  1,088 & 54.5 &  550& 51.4 &  7,946 & 62.6 \\ 

\bottomrule
\end{tabular}
\label{tab:random_testcase}
\end{table}

%% file: tables/testing_result.tex
\begin{table*}[]
\caption{\small{The  results of differential testing. "CPU Time" denotes the sum of the time used by CPU for all the test cases, which is in seconds. We do not count the overall CPU time for real devices as different devices have different CPUs.
"Inst": Instructions; "Enc": Encodings; "Cons\_Unpre": Constraint UNPREDICTABLE; "Annotation\_Def": Defined in Annotation;  $\langle X \mid Y\rangle$: For each instruction set and overall result, X denotes the number of the attribute indicated by the row name  while Y denotes the percentage of dividing X by Z. For data in "Testing Result", Z stands for the row "Generated Instruction Streams", "All Instruction Encodings", and "All Instructions". For data in "Inconsistent Behaviors" and "Root Cause", Z stands for "Inconsistent Instruction Streams". $[M \mid N]$: For each instruction set and overall result, M denotes the number of instruction encodings while N denotes how many instructions M belongs to.
We do not calculate the percentage of instruction encodings or instructions for "Inconsistent Behaviors" and "Root Cause" as one encoding or instruction can have more than one inconsistent behaviors, resulting from different root causes.}}

\footnotesize
\scalebox{0.7}{
\begin{tabular}{l|c|c|cc|c|c}

 \toprule
Arachitecture                   & ARMv5            & ARMv6   & \multicolumn{2}{c|}{ARMv7-a}     & ARMv8-a      & Overall \\ \hline\hline
Experiment Setup  &  &  &   &  & \\ \hline
QEMU Binary                   &   qemu-arm               &   qemu-arm       & \multicolumn{2}{c|}{qemu-arm   }          &    qemu-aarch64           &    -     \\ 
QEMU Model              & ARM926           & ARM1176 & \multicolumn{2}{c|}{Cortex-A7} & Cortex-A72 &    -     \\ 
Device Name & OLinuXino IMX233 & RaspberryPi  Zero & \multicolumn{2}{c|}{RaspberryPi 2B} & Hikey 970  & - \\ 
Device Model    & ARM926           & ARM1176           & \multicolumn{2}{c|}{Cortex-A7}      & Cortex-A73 & - \\ 
Instruction Set                 & A32              & A32     & A32           & T32\&T16       & A64        &    -     \\ 
\rowcolor[HTML]{C0C0C0} 
Generated Instruction Streams                       & 870,221 & 870,221& 870,221       & 809,728        & 1,094,700  &    2,774,649     \\
\rowcolor[HTML]{C0C0C0} 
All Instruction Encodings     & 550 & 550& 550       & 609      & 839  &    1,998     \\ 
\rowcolor[HTML]{C0C0C0} 
All Instructions    & 481 & 481& 481       & 462      & 581  &    1,070     \\ 
CPU Time (QEMU)  & 530.5 &540.6 & 538.2 &462.1&625.9 & 2697.3\\
CPU Time (Device) &46238.0 & 6901.7 & 6194.2 & 5276.0 &9145.0 & -\\
\hline
Testing Result  &  \multicolumn{5}{c}{The percentage is based on the number of generated instructions and all encodings} & \\ \hline \hline
 Read/Write to SP/FP      &  $\langle 37,002 \mid 4.3\%\rangle$            &  $\langle 36,879 \mid 4.2\%\rangle$   &  $\langle 37,002 \mid 4.3\% \rangle$         &  $\langle 24,582 \mid 3.0\%\rangle$          &  $\langle 45,985 \mid 4.2\%\rangle$      &    $\langle 107,569 \mid 3.9\%\rangle$       \\

 Branch       &  $\langle 3,557 \mid 0.4\%\rangle$            &  $\langle 5,431 \mid 0.6\%\rangle$    &  $\langle 1,821 \mid 0.2\%\rangle$          &  $\langle 1,048 \mid 0.1\%\rangle$           &  $\langle 210 \mid 0.0\%\rangle$         &    $\langle 7,017 \mid 0.4\%\rangle$      \\

Consistent Instruction Streams   &  $\langle 794,418 \mid 91.3\%\rangle$          & $\langle 818,744 \mid 94.1\%\rangle$  &  $\langle 773,906 \mid 88.9\%\rangle$        &  $\langle 738,369 \mid 91.2\%\rangle$         &  $\langle 1,031,901 \mid 94.3\%\rangle$   &       $\langle  2,598,695 \mid 93.7\%\rangle$    \\ 
\rowcolor[HTML]{C0C0C0} 
 Inconsistent Instruction Streams &  $\langle 35,244 \mid 4.1\%\rangle$           &  $\langle 9,167 \mid 1.1\%\rangle$    &  $\langle 57,492 \mid 6.6\% \rangle$         &  $\langle 45,729 \mid 5.6\%\rangle$          &  $\langle 16,604 \mid 1.5\% \rangle$      &    $\langle 155,642 \mid 5.6\%\rangle$      \\ 
\rowcolor[HTML]{C0C0C0} 
Inconsistent Instruction Encodings &           $\langle 108 \mid 19.6\%\rangle$        &   $\langle 86 \mid 15.6\%\rangle$        &        $\langle 270 \mid 49.0\%\rangle$         &        $\langle 266 \mid 43.6\%\rangle$          &     $\langle 15 \mid 1.8\%\rangle$         &  $\langle 600 \mid 30\%\rangle$  \\
\rowcolor[HTML]{C0C0C0} 
Inconsistent Instructions &           $\langle 98 \mid 20.4\%\rangle$        &   $\langle 84 \mid 17.5\%\rangle$        &        $\langle 229 \mid 47.6\%\rangle$         &        $\langle 223 \mid 48.3\%\rangle$          &     $\langle 13 \mid 2.2\%\rangle$         &  $\langle 511 \mid 47.8\%\rangle$  \\
\hline \hline 
Inconsistent Behaviors &  \multicolumn{5}{c}{The percentage is based on the number of inconsistent instructions}& \\ \hline
 $Sig_E \ne Sig_R \ne 0$  $\langle I\_S \mid I\_S\_R\rangle$  &  $\langle 10,748 \mid 30.6\%\rangle$  &  $\langle 6,853 \mid 74.8\%\rangle$   &  $\langle 35,628 \mid 62.0\%\rangle$   &  $\langle 24,094 \mid 52.7\%\rangle$   &  $\langle 667 \mid 4.0\%\rangle$  &   $\langle 73,619 \mid 47.3\%\rangle$ \\ 
 $Sig_E \ne Sig_R \ne 0$ $[Enc \mid Inst]$ & $[57 \mid 51]$   & $[63 \mid 61]$ & $[209 \mid 186]$  &$[212 \mid 181]$  & $[7 \mid 5]$&  $[458 \mid 401]$\\ 
 $Sig_E \ne Sig_R = 0$  $\langle I\_S \mid I\_S\_R\rangle$ &  $\langle 4,559 \mid 13.0\%\rangle$   &  $\langle 1,433 \mid 15.6\%\rangle$  &  $\langle 551 \mid 1.0\%\rangle$   &  $\langle 836 \mid 1.8\%\rangle$  &  $\langle 0 \mid 0.0\%\rangle$ &  $\langle 7,197 \mid 4.6\%\rangle$ \\ 
 $Sig_E \ne Sig_R = 0$ $[Enc \mid Inst]$& $[34 \mid 33]$  & $[47 \mid 46]$ &  $[18 \mid 18]$ &  $[9 \mid 0]$&$[0 \mid 0]$  & $[72 \mid 70]$\\ 
 $Sig_R \ne Sig_E  = 0$  $\langle I\_S \mid I\_S\_R\rangle$ &  $\langle 13,827 \mid 39.3\%\rangle$  &  $\langle 3 \mid 0.0\%\rangle$  &   $\langle 21,135 \mid 36.8\%\rangle$  &  $\langle 19,915 \mid 43.6\%\rangle$  &  $\langle 11,219 \mid 67.6\%\rangle$ &  $\langle 65,599 \mid 42.1\%\rangle$ \\ 
 $Sig_R \ne Sig_E  = 0$ $[Enc \mid Inst]$& $[42 \mid 35]$ & $[1 \mid 1]$  & $[193 \mid 168]$  & $[203 \mid 177]$ & $[7 \mid 7]$ & $[420 \mid 367]$\\ 
$Sig_R = Sig_E  \ne 0$  $\langle I\_S \mid I\_S\_R\rangle$ &  $\langle 0 \mid 0.0\%\rangle$  &  $\langle 67 \mid 0.7\%\rangle$  &   $\langle 174 \mid 0.3\%\rangle$  &  $\langle 873 \mid 1.9\%\rangle$  &  $\langle 4,716 \mid 28.4\%\rangle$  &  $\langle 5,763 \mid 3.7\%\rangle$ \\
$Sig_R = Sig_E  \ne 0$ $[Enc \mid Inst]$& $[0 \mid 0]$ & $[6 \mid 6]$ &   $[18 \mid 18]$ & $[18 \mid 15]$ &$[3 \mid 3]$ & $[39 \mid 36]$\\
  $Sig_R = Sig_E  = 0$  $\langle I\_S \mid I\_S\_R\rangle$ &  $\langle 2,389 \mid 6.8\% \rangle$   &  $\langle 318 \mid 3.5\% \rangle$  &   $\langle 3 \mid 0.0\%\rangle$  &  $\langle 8 \mid 0.0\%\rangle$  &  $\langle 0 \mid 0.0\%\rangle$  &  $\langle 2,410 \mid 1.5\%\rangle$  \\
  $Sig_R = Sig_E  = 0$ $[Enc \mid Inst]$& $[25 \mid 25]$  & $[6 \mid 6]$ &  $[1 \mid 1]$ &  $[1 \mid 1]$& $[0 \mid 0]$& $[27 \mid 27]$\\
  Others  $\langle I\_S \mid I\_S\_R\rangle$  &  $\langle 3,721 \mid 10.6\%\rangle$   &  $\langle 493 \mid 5.4\%\rangle$  &  $\langle 1 \mid 0.0\%\rangle$   &  $\langle 3 \mid 0.0\%\rangle$  &  $\langle 2 \mid 0.0\%\rangle$ &  $\langle 4,218 \mid 2.8\%\rangle$ \\ 
    Others $[Enc \mid Inst]$ & $[8 \mid 8]$& $[2 \mid 2]$ & $[1 \mid 1]$  & $[3 \mid 2]$ & $[2 \mid 2]$ & $[14 \mid 13]$\\ \hline \hline 
  Root Cause & \multicolumn{5}{c}{The percentage is based on the number of inconsistent instructions} &\\\hline 
  Bugs of QEMU  $\langle I\_S \mid I\_S\_R\rangle$ &  $\langle 1 \mid 0.0\%\rangle$   &  $\langle 1 \mid 0.0\%\rangle$ &    $\langle 1 \mid 0.0\%\rangle$  &  $\langle 583 \mid 1.3\%\rangle$  &  $\langle 2 \mid 0.0\%\rangle$ &  $\langle 586 \mid 0.4\%\rangle$ \\
Bugs of QEMU $[Enc \mid Inst]$& $[1 \mid 1]$ & $[1 \mid 1]$ &  $[1 \mid 1]$ & $[10 \mid 7]$ &$[2 \mid 2]$ & $[13 \mid 10]$\\
  UNPREDICTABLE  $\langle I\_S \mid I\_S\_R\rangle$  &  $\langle 35,147 \mid 99.7\%\rangle$  &  $\langle 8,913 \mid 97.2\%\rangle$  &   $\langle 57,428 \mid 99.9\%\rangle$  &  $\langle 44,869 \mid 98.1\%\rangle$   &  $\langle 2 \mid 0.0\%\rangle$  & $\langle 137,828 \mid 88.6\%\rangle$ \\
    UNPREDICTABLE  $[Enc \mid Inst]$& $[106 \mid 96]$ & $[77 \mid 75]$ &  $[265 \mid 224]$ & $[261 \mid 218]$ & $[0 \mid 0]$ & $[570 \mid 483]$\\
  Cons\_UNPRE  $\langle I\_S \mid I\_S\_R\rangle$ &  $\langle 0 \mid 0.0\%\rangle$  &  $\langle 0 \mid 0.0\%\rangle$  &  $\langle 0 \mid 0.0\%\rangle$ & $\langle 0 \mid 0.0\%\rangle$  &  $\langle 16,602 \mid 100.0\%\rangle$ &  $\langle 16,602 \mid 10.7\%\rangle$ \\
    Cons\_UNPRE $[Enc \mid Inst]$& $[0 \mid 0]$ & $[0 \mid 0]$&  $[0 \mid 0]$ & $[0 \mid 0]$& $[15 \mid 13]$ & $[15 \mid 13]$\\
  Annotation\_Def  $\langle I\_S \mid I\_S\_R\rangle$  &  $\langle 0 \mid 0.0\%\rangle$   &  $\langle 253 \mid 2.8\% \rangle$  &  $\langle 63 \mid 0.1\%\rangle$   &   $\langle 277 \mid 0.6\%\rangle$  &  $\langle 0 \mid 0.0\%\rangle$ &  $\langle 530 \mid 0.3\%\rangle$ \\
  Annotation\_Def $[Enc \mid Inst]$& $[0 \mid 0]$ &$[8 \mid 8]$  &  $[4 \mid 4]$ & $[4 \mid 4]$  & $[0 \mid 0]$& $[12 \mid 12]$\\
  Others  $\langle I\_S \mid I\_S\_R\rangle$ &  $\langle 96 \mid 0.3\%\rangle$  &  $\langle 0 \mid 0.0\%\rangle$  &  $\langle 0 \mid 0.0\%\rangle$   &   $\langle 0 \mid 0.0\%\rangle$ &  $\langle 0 \mid 0.0\%\rangle$ &  $\langle 96 \mid 0.0\%\rangle$ \\
  Others $[Enc \mid Inst]$ & $[1 \mid 1]$ & $[0 \mid 0]$ &  $[0 \mid 0]$ & $[0 \mid 0]$  & $[0 \mid 0]$ & $[1 \mid 1]$\\
  
  \bottomrule
\end{tabular}}
\label{tab:testing_result}
\end{table*}

%% file: tables/android.tex
\begin{table}[]
\caption{\small{The statistics on detecting emulators}}
\centering
\footnotesize

\begin{tabular}{ccccc}
\toprule
Mobile Type       & CPU            & A64 & A32 & T32 \& T16 \\ \hline \hline
Samsung S8        & SnapDragon 835 &  \checkmark    &   \checkmark   &  \checkmark          \\ 
Huawei Mate20     & Kirin 980      &  \checkmark    &   \checkmark   &   \checkmark          \\
IQOO Neo5         & SnapDragon 870 &  \checkmark    &  \checkmark    &  \checkmark          \\ 
Huawei P40        & Kirin 990      &  \checkmark    &  \checkmark    &  \checkmark           \\ 
Huawei Mate40 Pro & Kirin 9000     &  \checkmark    & \checkmark     &    \checkmark          \\ 
Honor 9           & Kirin 960      & \checkmark     &   \checkmark   &     \checkmark        \\ 
Honor 20          & Kirin 710      & \checkmark     &  \checkmark    &  \checkmark           \\ 
Blackberry Key2   & SnapDragon 660 &  \checkmark    &  \checkmark    & \checkmark            \\ 
Google Pixel      & SnapDragon 821 &   \checkmark   &   \checkmark   &   \checkmark          \\ 
Samsung Zflip     & SnapDragon 855 &  \checkmark    &    \checkmark  &   \checkmark          \\ 
Google Pixel3     & SnapDragon 845 &  \checkmark    &    \checkmark  &  \checkmark           \\ \bottomrule
\end{tabular}
 
\label{tab:android}
\end{table}

%% file: 06_discussion.tex
\section{Discussion}
\label{sec:discussion}

\smallskip
\noindent 
\textbf{Testing Instructions in Privileged Environments}\tab
Currently, the generated instruction streams are tested under unprivileged mode in both CPU emulators and real devices. Some instruction streams may have different execution results under privileged mode. For instance, instruction \code{WFI}, which results in a bug of QEMU user-mode, may not be an inconsistent instruction while executing in privileged mode. We plan to port \sysname to kernel-space for a more thorough testing.

\smallskip
\noindent 
\textbf{Testing Instruction Stream Sequences}\tab
\sysname now tests only one instruction stream each time during the differential testing. We can also test multiple instruction streams (instruction stream sequences) in the differential testing. The instruction stream sequences may trigger multiple system states and can test the decoding/executing logic of real devices and CPU emulators towards different state flags. How to design representative instruction stream sequences, and how to locate the inconsistent one will be the challenge. This will be our future work. Nevertheless,
We have already discovered a huge number of inconsistent instruction streams with \sysname, covering 47.8\% of the whole instructions. Every instruction stream sequence that contain the inconsistent instruction stream can result in inconsistent behaviors.

\smallskip
\noindent 
\textbf{Detecting (Ab)Used Inconsistent Instructions}\tab
The experi\-me\-nts in Section~\ref{sec:anti_emu} and \ref{sec:anti_fuzz} show that attackers or vendors can (ab)use these inconsistent instructions to prevent the released binary from being monitored or fuzzed. It is not easy to recognize these inconsistent instructions due to the huge number of inconsistent instruction streams from 511 instructions. Some of these instructions are even commonly used (i.e., STR and BLX instruction). 
Apart from this, attackers can encrypt these instruction streams as data. Then these encrypted instruction streams can be decrypted and executed during runtime, which can increase the bar for detection.
Thus, it is challenging to precisely detect those inconsistent instruction streams if they are used to conduct anti-emulation or anti-fuzzing tasks. 
Furthermore, how to hide these inconsistent instruction streams from being detected is a \textit{Cat and Mouse} problem. Stealthily using these instructions is not the purpose of our work.

\smallskip
\noindent 
\textbf{Other Emulators and Architectures}\tab
We test QEMU as it is one of the state-of-the-art emulators, which is well maintained and used by many industry tools (e.g., Android Studio) and academic prototypes~\cite{davanian2019decaf++,henderson2014decaf,yan2012droidscope,alwabel2014safe,wei2015mose,carmony2016extract, chipounov2011s2e,jiang2010stealthy, chen2016firmadyne,feng2014mace,luo2016repackage}. However, \sysname can also be used to test the other emulators. Generally, all the tools (e.g., angr, Unicorn) including an emulated execution engine can be tested, which will be left as the future work. 
The whole framework  of \sysname is architecture independent. However, we rely on  ARM ASL to generate the test cases, which can explore multiple behaviors. 
If other architectures propose such kinds of specification language, we are able to generate the test cases. Otherwise, new test case generation algorithm should be developed.

%% file: 07_related_work.tex
\section{Related Work}
\subsection{Testing CPU Emulators}
Several works are proposed to test the CPU emulators.
Lorenzo et al. proposed EmuFuzzer to test the CPU emulators~\cite{martignoni2009testing,martignoni2013methodology}. However, the seed used for testing mainly relies on randomization and a CPU-assisted mechanism, which may not cover all the CPU behaviors. 
Apart from testing user-level instructions, KEmuFuzzer is proposed to test the whole system emulators~\cite{martignoni2010testing}. However, KEmuFuzzer relies on the manually written template to generate test cases. For better test case coverage, PokeEMU~\cite{lorenzo2012} is proposed. PokeEMU utilizes binary  symbolic execution to generate more test cases from a high-fidelity emulator and apply these test cases on low-fidelity emulators. However, whether the high-fidelity emulator strictly follow the rule of specification is unknown. Furthermore, all the above mentioned works target on x86/x64 architectures. With the development of embedded systems and mobiles, the faithfully emulating ability for ARM architecture is a urgent need. Our work targets on ARM architecture and generates test cases from the specification itself (i.e., ARM ASL). The evaluation results show that we can find the real bugs and many inconsistent implementations between real devices and emulators, which can be abused by attackers.

\subsection{Differential Testing}
Differential testing is introduced by McKeeman et al.~\cite{mckeeman1998differential} to detect the implementation bugs by comparing the inconsistent behaviors between different software. For example, Yang et al. proposed Csmith, a powerful tool that can generate multiple C programs. With Csmith, hundreds of bugs are detected in the C compiler. Regarding the same goal, Le et al. introduced equivalence modulo
inputs (EMI)~\cite{le2014compiler} and many other differential testing tools are built based on EMI to validate the compiler implementations~\cite{lidbury2015many,sun2016finding}. 

Apart from testing compilers, researchers also utilize differential testing to validate the Database Management Systems (DBMS). Slutz et al. proposed the tool RAGS to explore bugs by executing
different SQL queries on multiple DBMS. Though it is effective, it can only support a small set of SQL statements. Gu et al. evaluate the accuracy of DBMS optimizer by using options and hints to force the generation of different query plans. Jung et al. developed APOLLO~\cite{jung2019apollo} to test the performance regression bugs in DBMSs .

Furthermore, differential testing is powerful and  applied to different domains such as testing SMT solvers~\cite{winterer2020validating,winterer2020unusual}, JVM implementations~\cite{kapus2017automatic} , symbolic execution engines~\cite{kapus2017automatic}, and PDF readers~\cite{kuchta2018correctness}.
.

\subsection{Anti-Emulation Technique}
Previous anti-emulation works~\cite{raffetseder2007detecting} divide the anti-emulation technique into three categories. They are differences in behavior, differences in timing, and hardware specific values. 
Our work can automatically locate the inconsistent instructions, which result in different behavior and can be used by the previous anti-emulation technique. Jang et al.~\cite{jang2019rethinking} address the importance of anti-emulation techniques on protecting the Commercial-Off-the-Shelf (COTS) software from being debugged or used without buying hardwares. They propose three different anti-emulation techniques. However, some techniques rely on the race condition and are not easy to trigger.

%% file: 08_conclusion.tex
\section{Conclusion}
We design and implement \sysname, a framework that can automatically locate the inconsistent ARM instructions, which can result in inconsistent behaviors between CPU emulator and real devices. With \sysname, we generate 2,774,649 representative instruction streams and detect 155,642 inconsistent instruction streams covering 30\% of the  instruction encodings and 47.8\% instructions. By analyzing the root cause of inconsistent instructions, we noticed four bugs of QEMU, which are confirmed by QEMU developers, covering 13 instruction encodings including very commonly used ones (e.g., \code{STR}, \code{BLX}). We also demonstrate the capability of inconsistent instructions on detecting emulators, anti-emulation, and anti-fuzzing.